\documentclass[usenatbib]{mn2e}
\usepackage{graphicx}
\usepackage{epsfig}
\include{epsf}
\newcommand\aj{AJ}
\newcommand\apj{ApJ}
\newcommand\mnras{MNRAS}

\title[Quasar SZ Simulations]{Simulations of the Sunyaev-Zeldovich Effect from Quasars}
\author[Chatterjee et al.]{Suchetana Chatterjee$^{1}$ Tiziana Di Matteo$^{2}$, Arthur Kosowsky$^{1}$ \& Inti Pelupessy$^{2}$\\
$^{1}$Department of Physics and Astronomy, University of Pittsburgh, Pittsburgh, PA 15260 USA\\
$^{2}$McWilliam's Center for Cosmology, Carnegie Mellon University, Pittsburgh, PA 15213 USA}
\begin{document}



\maketitle

\label{firstpage}

\begin{abstract}
Quasar feedback has most likely a substantial but only partially
understood impact on the formation of structure in the universe. A potential direct probe of this
feedback mechanism is the Sunyaev-Zeldovich effect: energy emitted from quasar heats
the surrounding intergalactic medium and induce a distortion in the microwave background radiation passing through the region. Here we examine the formation of such hot quasar bubbles
using a cosmological hydrodynamic simulation which includes a self-consistent treatment of black hole 
growth and associated feedback, along with radiative gas cooling and star formation.
From this simulation, we construct microwave maps of the resulting Sunyaev-Zeldovich effect around black holes with a range of masses and redshifts. The size of
the temperature distortion scales approximately with black hole mass and accretion rate, with a typical
amplitude up to a few micro-Kelvin on angular scales around 10 arcseconds. We discuss
prospects for the direct detection of this signal with current and future single-dish
and interferometric observations, including ALMA and CCAT. These measurements will be
challenging, but will allow us to characterize the evolution and growth of supermassive black holes and the role of their energy feedback on galaxy formation.
\end{abstract}


\section{Introduction}

The temperature fluctuations in the cosmic microwave background, as measured
by the Wilkinson Microwave Anisotropy Probe (WMAP) satellite (Bennett et al.\
2003) and numerous other microwave experiments (e.g., Dawson et al.\ 2002; Rajguru et al.\ 2005; Reichardt et al. 2008) have proven to be
the single most powerful tool in constraining cosmology (Spergel et al.\
2007). The temperature anisotropy has been mapped with large statistical
significance on angular scales down to around a quarter degree, where the
dominant physical mechanisms contributing to the fluctuations arise from
density perturbations at the epoch of recombination. Attention is now turning
to arcminute angular scales, where temperature fluctuations arise due to
interaction of the microwave photons with matter in the low-redshift universe
(for a brief review, see Kosowsky 2003). These low-redshift and small-angle
anisotropies are collectively known as ``secondary anisotropies'' in the
microwave background. The most prominent among them is the Sunyaev-Zeldovich
(SZ) effect (Sunyaev \& Zeldovich 1972) from the inverse Compton scattering of
the microwave photons due to hot electrons. The SZ effect provides a powerful
method for detecting accumulations of hot gas in the universe (Carlstrom Holder \& Reese 2002).  Galaxy clusters,
which contain the majority of the thermal energy in the universe, provide the
largest SZ signal; clusters were first detected this way through pioneering
measurements over the past decade (e.g, Marshall et al.\ 2001) and thousands
of them will be detected by the upcoming SZ surveys like the Atacama Cosmology
Telescope (ACT) (Kosowsky et al. 2006) and the South Pole Telescope (SPT)
(Ruhl et al. 2004). However, a number of other astrophysical processes will
also create SZ distortions. This includes SZ distortion from peculiar velocities during reionization (McQuinn et al. 2005, Illiev et al. 2006), supernova-driven galactic winds (Majumdar, Nath, \& Chiba 2001), black hole seeded proto-galaxies (Aghanim, Ballad \& Silk 2000), kinetic SZ from Lyman Break Galaxy outflow (Babich \& Loeb 2007), effervescent heating in groups and clusters of galaxies (Roychowdhury, Ruszkowski \& Nath 2005) and supernova from first generation of stars (Oh, Cooray, \& Kamionkowski 2003). The SZ distortion in galactic scales (hot proto galactic gas) have been studied by different authors (e.g, de Zotti et al. 2004, Rosa-Gonz'alez et al. 2004, Massardi et al. 2008 ). Here we investigate one generic class of SZ signals: the hot
bubble surrounding a quasar powered by a supermassive
black hole. Probing black hole energy feedback via SZ distortions is one
direct observational route to understanding the growth and evolution of
supermassive black holes and their role in structure formation.  Analytic
studies of this signal have been done by several authors (e.g., Natarajan \&
Sigurdsson 1999; Yamada, Sugiyama \& Silk 1999; Lapi, Cavaliere \& De Zotti
2003; Platania et al.\ 2002; Chatterjee \& Kosowsky 2007); the current
numerical work complements a similar study by Scannapieco Thacker and Couchman
2008.
\begin{table*}
\begin{tabular}[t]{c|c|c|c|c|c|c}
\hline
\multicolumn{1}{c|}{Run}&
\multicolumn{1}{c|}{Boxsize}&
\multicolumn{1}{c|}{$N_{P}$}&
\multicolumn{1}{c|}{$m_{DM}$}&
\multicolumn{1}{c|}{$m_{gas}$}&
\multicolumn{1}{c|}{$\epsilon$}&
\multicolumn{1}{c}{$z_{end}$}\\
\multicolumn{1}{c|}{}&
\multicolumn{1}{c|}{($h^{-1}$ Mpc)}&
\multicolumn{1}{c|}{}&
\multicolumn{1}{c|}{($h^{-1}M\odot$)}&
\multicolumn{1}{c|}{($h^{-1}M\odot$)}&
\multicolumn{1}{c|}{($h^{-1}$ kpc)}&
\multicolumn{1}{c}{}\\
\hline
 D4 & 33.75 & $2\times216^{3}$ & $2.75\times10^{8}$ & $4.24\times10^{7}$ & 6.25 & 0.00 \\ 
 D6 (BHCosmo) & 33.75 & $2\times486^{3}$ & $2.75\times10^{7}$ & $4.24\times10^{6}$ & 2.73 & 1.00 \\ 
\hline 
\end{tabular}
\caption{The numerical parameters in the simulation. For the current study we have used the low-resolution version because we have a matching simulation with no black holes; resolution effects are discussed in Sec.~4.3. $N_{p}$, $m_{DM}$, $m_{gas}$, $ \epsilon $ and $z_{end}$ are defined as the total number of particles, mass of the dark matter particles, mass of the gas particles, gravitational softening length and final redshift run respectively.}

\end{table*}

Analytic models and numerical simulations of galaxy cluster formation indicate
that the temperature and the X-ray luminosity relation should be related as
$L_{x} \simeq T^{2}$ in the absence of gas cooling and heating (Peterson \&
Fabian 2006 and references therein). Observations show instead that $L_{x}
\simeq T^{3}$ over the temperature range 2 to 8 kev with a wide dispersion at
lower temperature, and a possible flattening above (Markevitch 1998; Arnaud \&
Evrard 1999; Peterson \& Fabian 2006). The simplest explanation for this
result is that the gas had an additional heating of 2 to 3 keV per particle
(Wu, Fabian \& Nulsen 2000; Voit et al.\ 2003). Several nongravitational
heating sources have been discussed in this context (Peterson \& Fabian 2006;
Morandi Ettori \& Moscardini 2007); quasar feedback (e.g., Binney \& Tabor 1995;
Silk \& Rees 1998; Ciotti \& Ostriker 2001; Nath \& Roychowdhury 2002; Kaiser
\& Binney 2003; Nulsen et al. 2004) is perhaps the most realistic
possibility. The effect of this feedback mechanism on different scales of
structure formation have been addressed by several authors (e.g., Mo \& Mao
2002; Oh \& Benson 2003; Granato et al. 2004). The mechanism of quasar heating in cluster cores has been observationally motivated by studies from McNamara et
al. 2005, Voit \& Donahue 2005 and Sanderson, Ponman \& O'Sullivan 2006 (see
McNamara \& Nulsen 2007 for a recent review). The impact of this
nongravitational heating in galaxy groups, which have shallower potential
wells and thus smaller intrinsic thermal energy than galaxy clusters, can also
be substantial (Arnaud \& Evrard 1999; Helsdon \& Ponman 2000; Lapi, Cavaliere \& Menci 2005). Observational
efforts to detect the impact of this additional heating source in the context
of quasar feedback have been carried out using galaxy groups in the Sloan Digital
Sky Survey (SDSS) by Weinmann et al.\ 2006, and with a Chandra group sample by
Sanderson, Ponman \& O'Sullivan 2006. Detailed theoretical studies of galaxy
groups using simulations which include quasar feedback have been undertaken by,
e.g., Zanni et al.\ 2005, Sijacki et al.\ 2007, and Bhattacharya, Di Matteo \&
Kosowsky 2007. At smaller scales, the impact of quasar feedback has been
investigated by Schawinski et al.\ 2007 with early-type galaxies in SDSS, and
has also been studied in several theoretical models of galaxy evolution (e.g,
Kawata \& Gibson 2005; Bower et al. 2006; Cattaneo et al.\ 2007).

Growing observational evidence points to a close connection between the
formation and evolution of galaxies, their central supermassive black holes
(e.g., Magorrian et al.\ 1998, Ferrarese \& Merritt 2000, Tremaine et al.\
2002) and their host dark matter halos (Merritt \& Ferrarese 2001; Tremaine et
al.\ 2002).  Several different groups have now investigated black hole growth
and the effects of quasar feedback in the cosmological context (e.g., Scannapieco
\& Oh 2004; Di Matteo, Springel \& Hernquist 2005; Lapi et al. 2006; Croton et al.\ 2006;
Thacker, Scannapieco \& Couchman 2006, Sijacki et al.\ 2007). Recently Di
Matteo et al.\ (2008) carried out a hydrodynamic cosmological simulation
following in detail the growth of supermassive black holes, using a simple but
realistic model of gas accretion and associated feedback. Using this simulation, we construct maps of the SZ distortion around black
holes of different masses at various redshifts. We demonstrate that the SZ
signal around quasars scales with black hole mass and accretion rate. A similar
approach has been taken by Scannapieco, Thacker \& Couchman (2008) from a
cosmological simulation with a different implementation of quasar feedback.
Predictions for the SZ distortion due to a phenomenological treatment of
galactic winds from simulations have been obtained previously by White,
Hernquist \& Springel (2002).

\begin{figure*}
  \begin{center}
    \begin{tabular}{cc}
      
\resizebox{55mm}{!}{\includegraphics{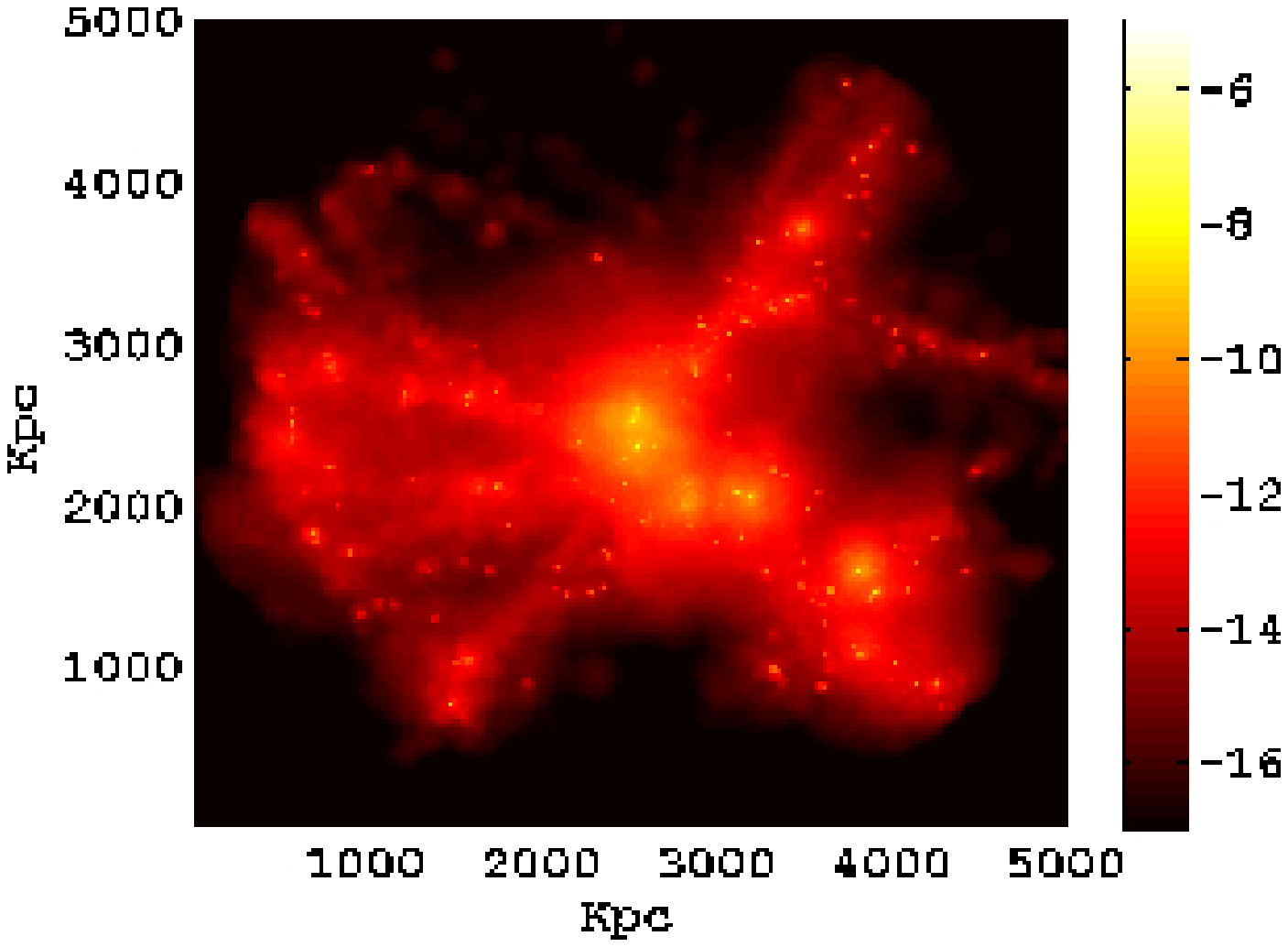}}
       \resizebox{55mm}{!}{\includegraphics{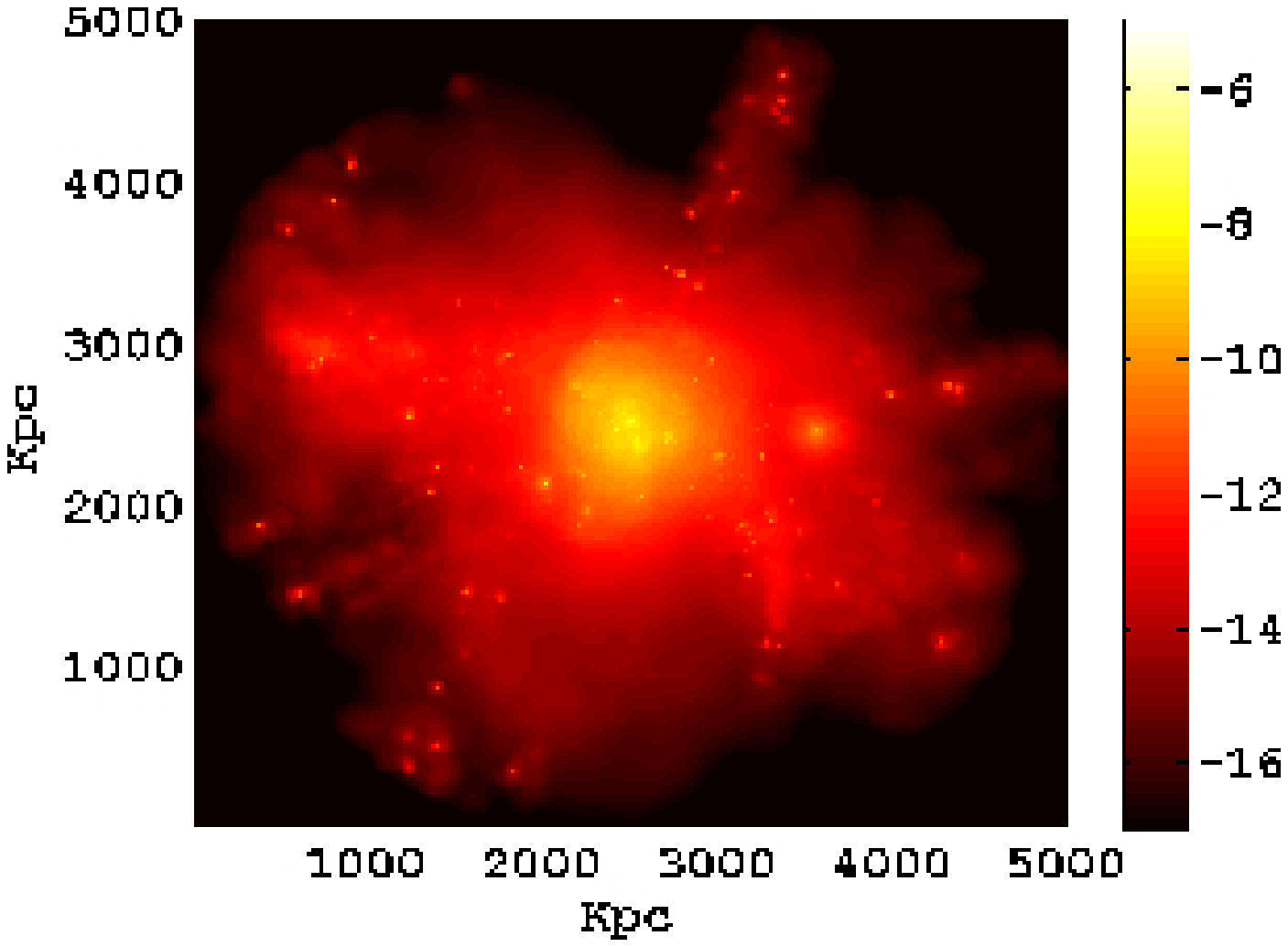}}
       \resizebox{55mm}{!}{\includegraphics{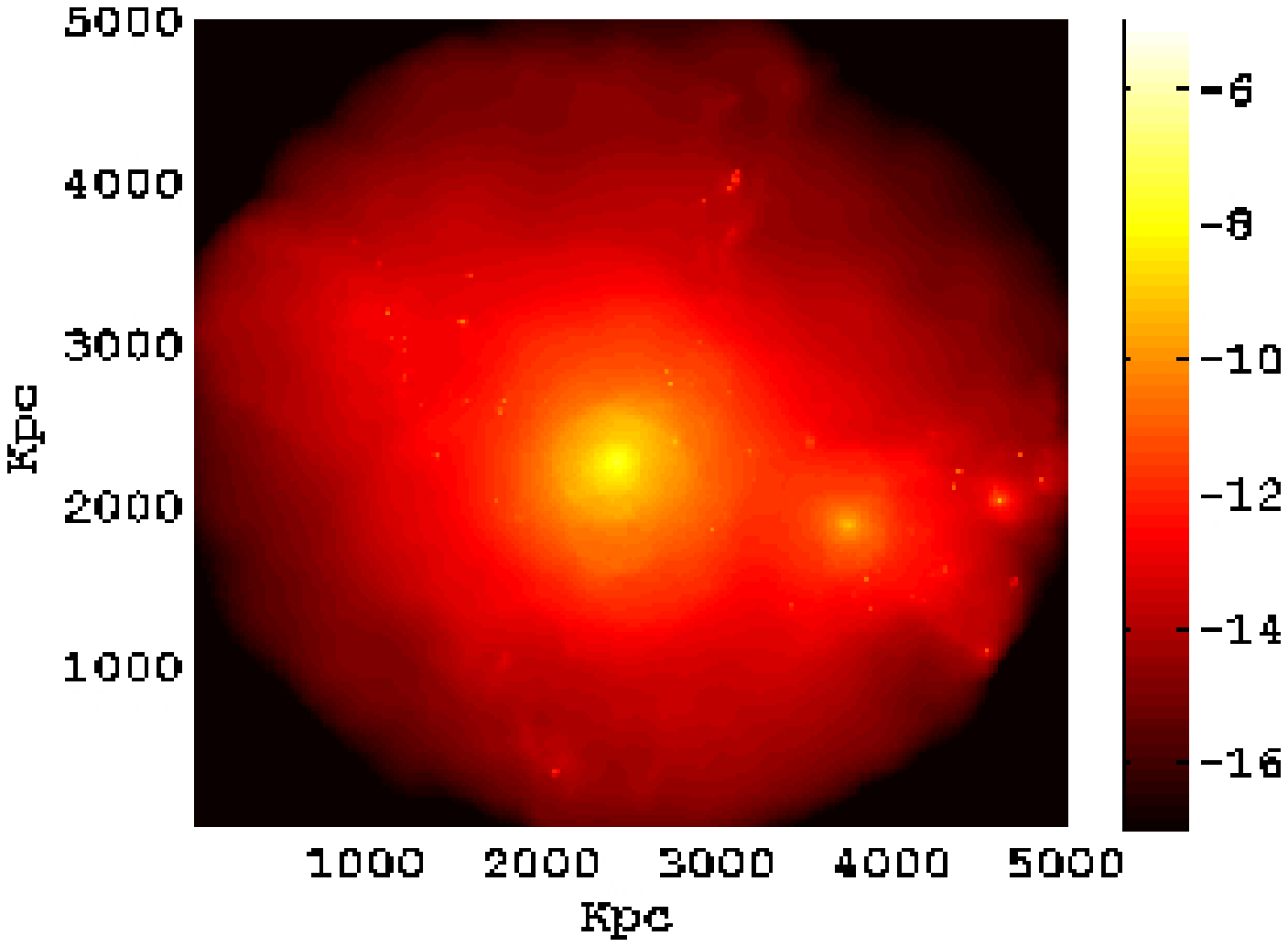}}\\
\resizebox{55mm}{!}{\includegraphics{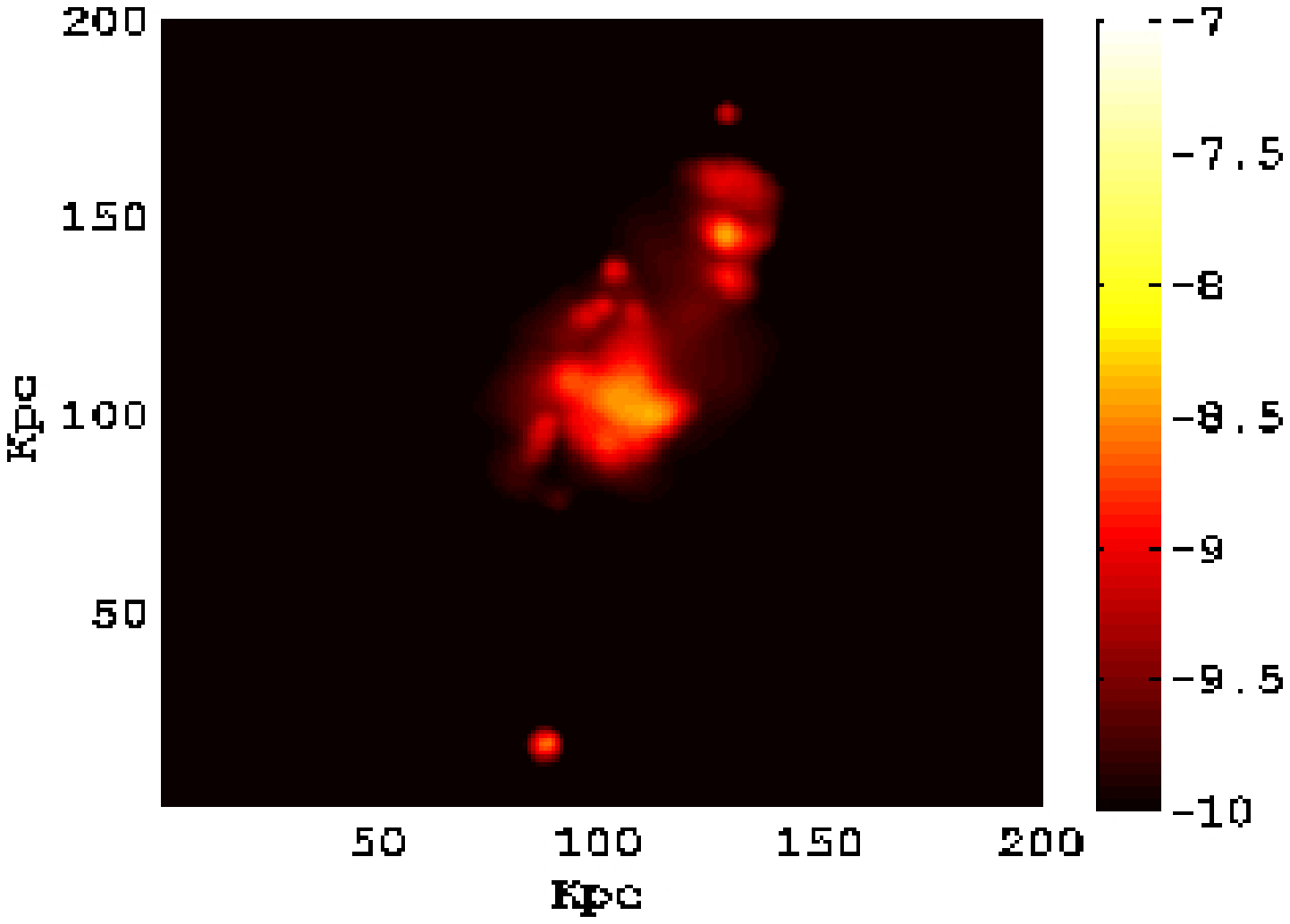}}
       \resizebox{55mm}{!}{\includegraphics{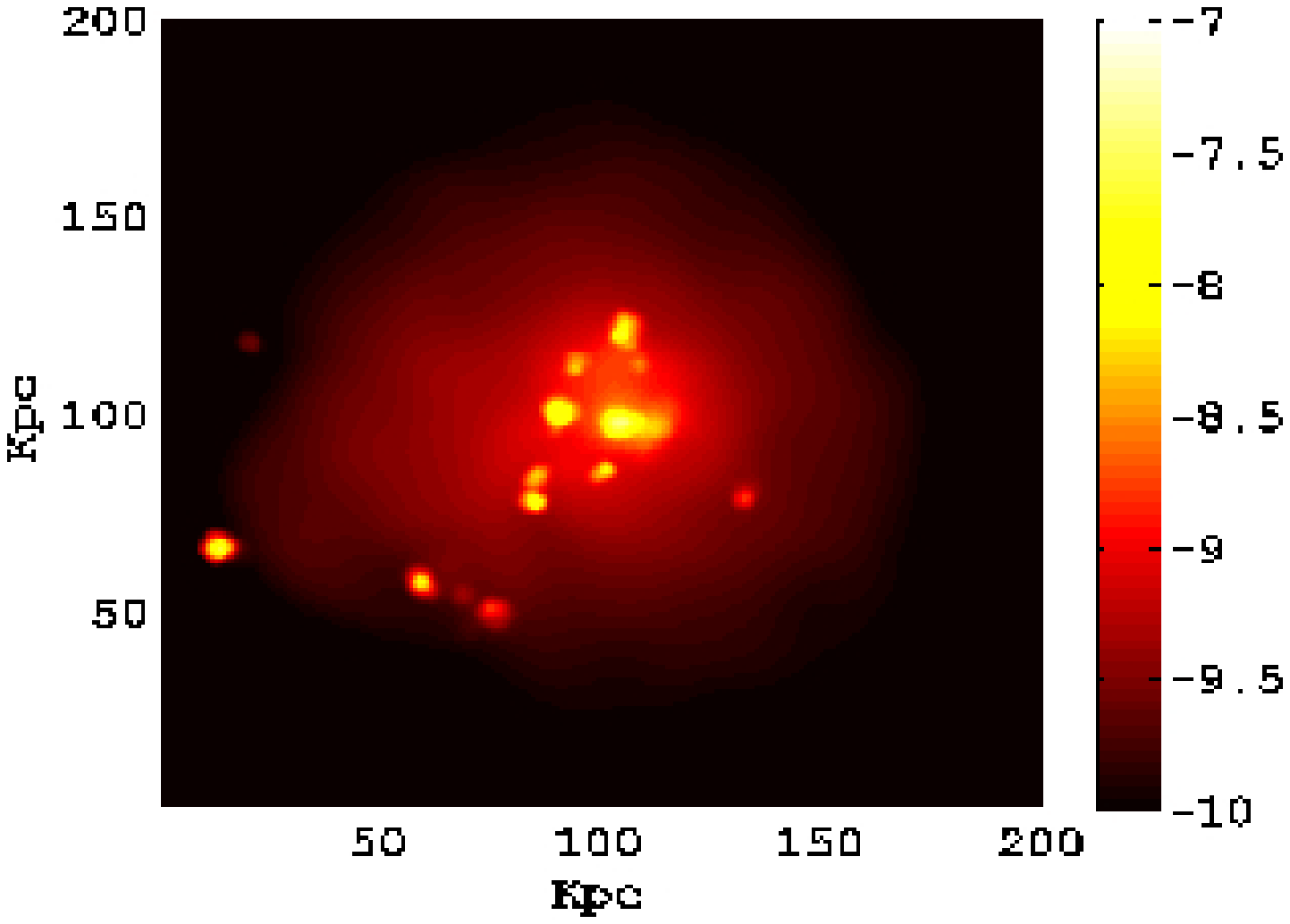}}
       \resizebox{55mm}{!}{\includegraphics{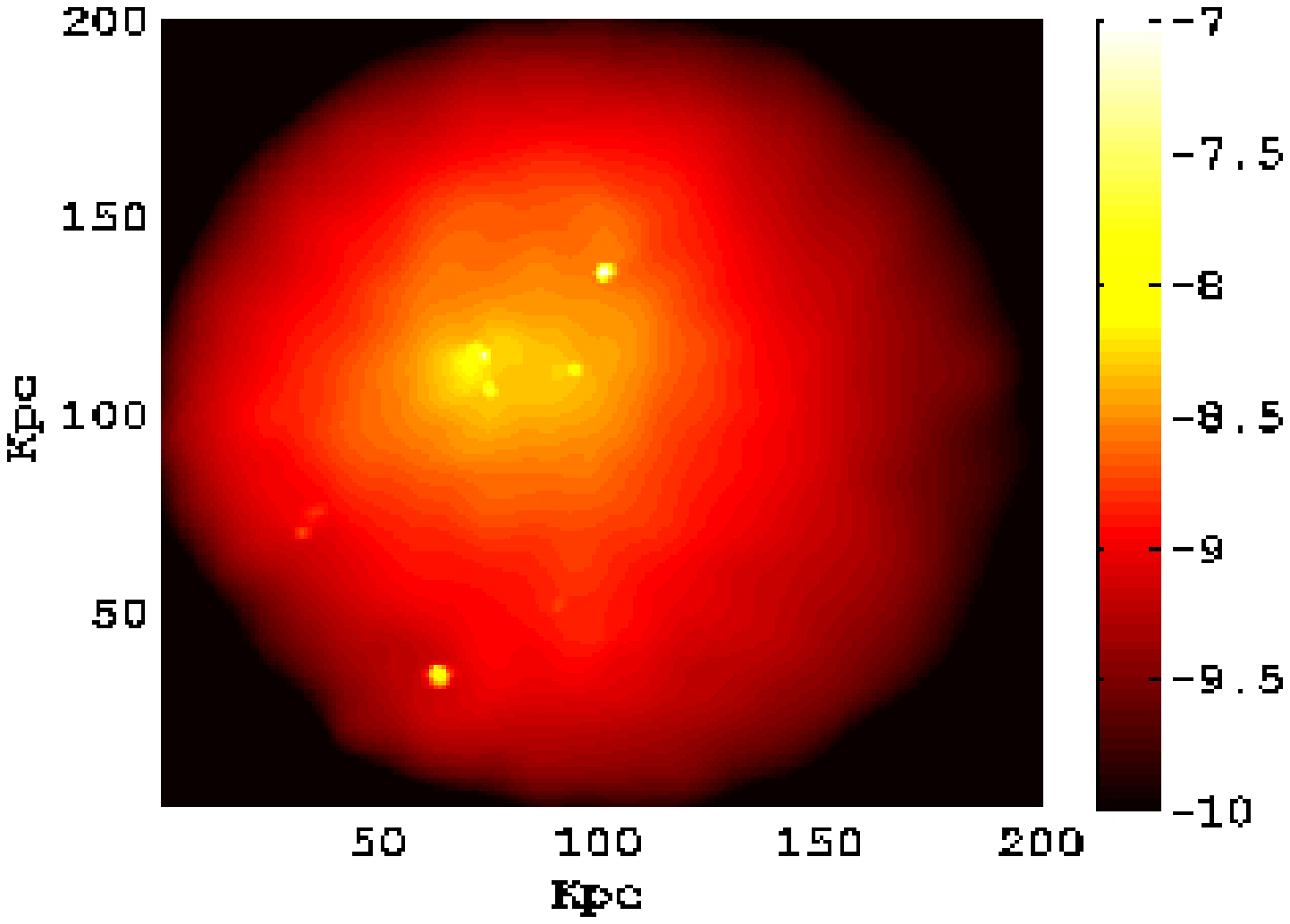}}\\

       \resizebox{55mm}{!}{\includegraphics{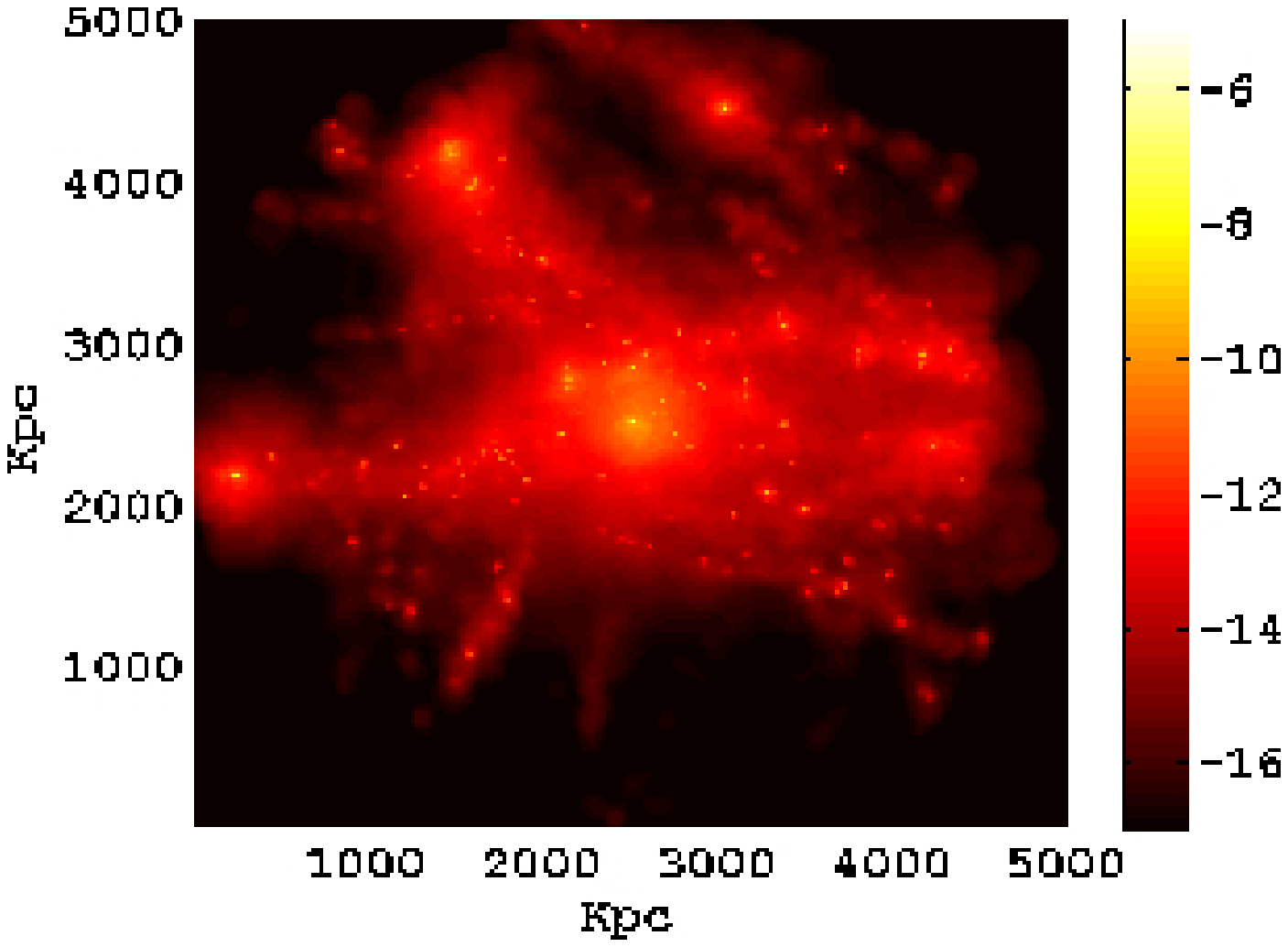}}
       \resizebox{55mm}{!}{\includegraphics{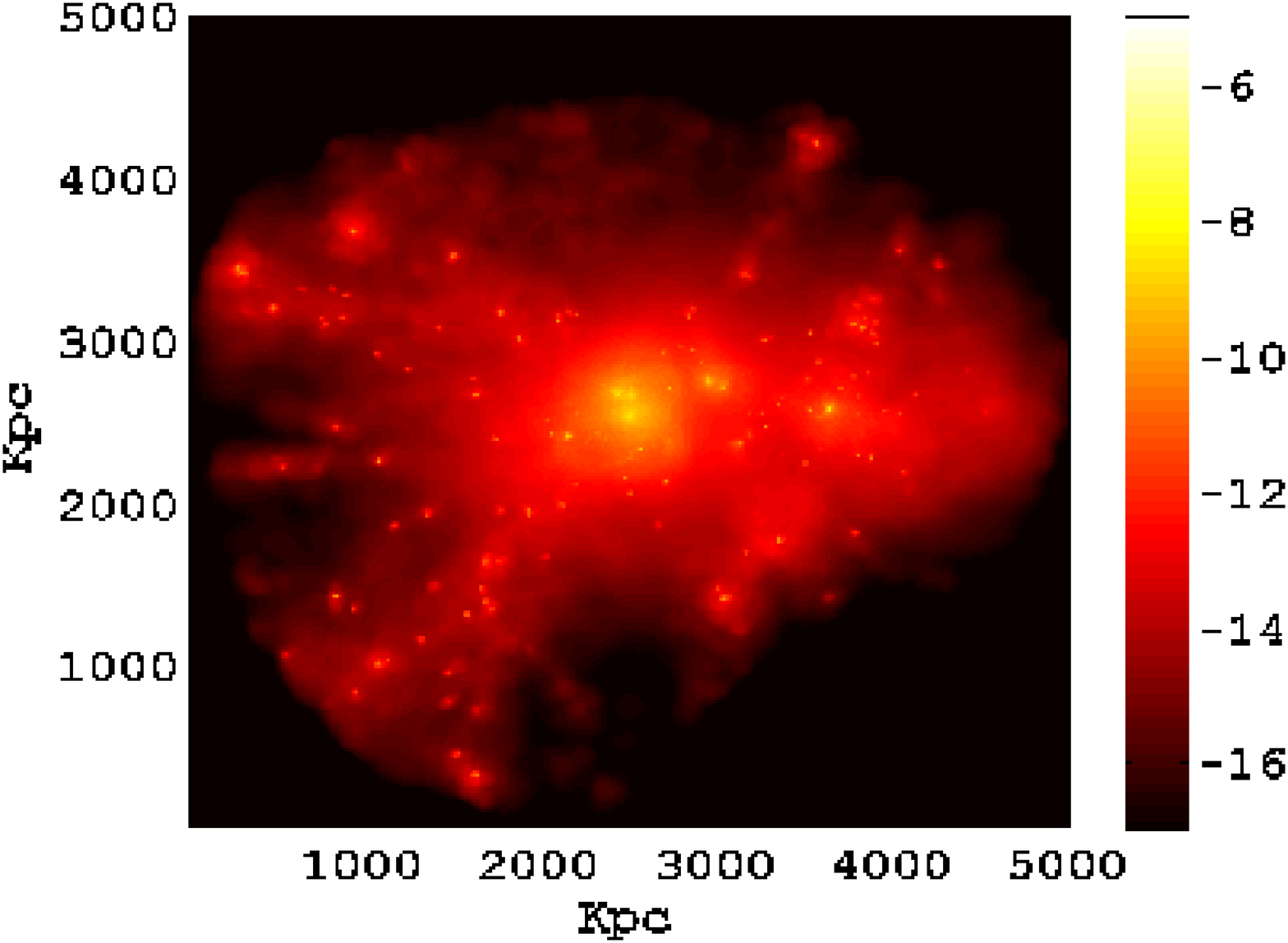}}
       \resizebox{55mm}{!}{\includegraphics{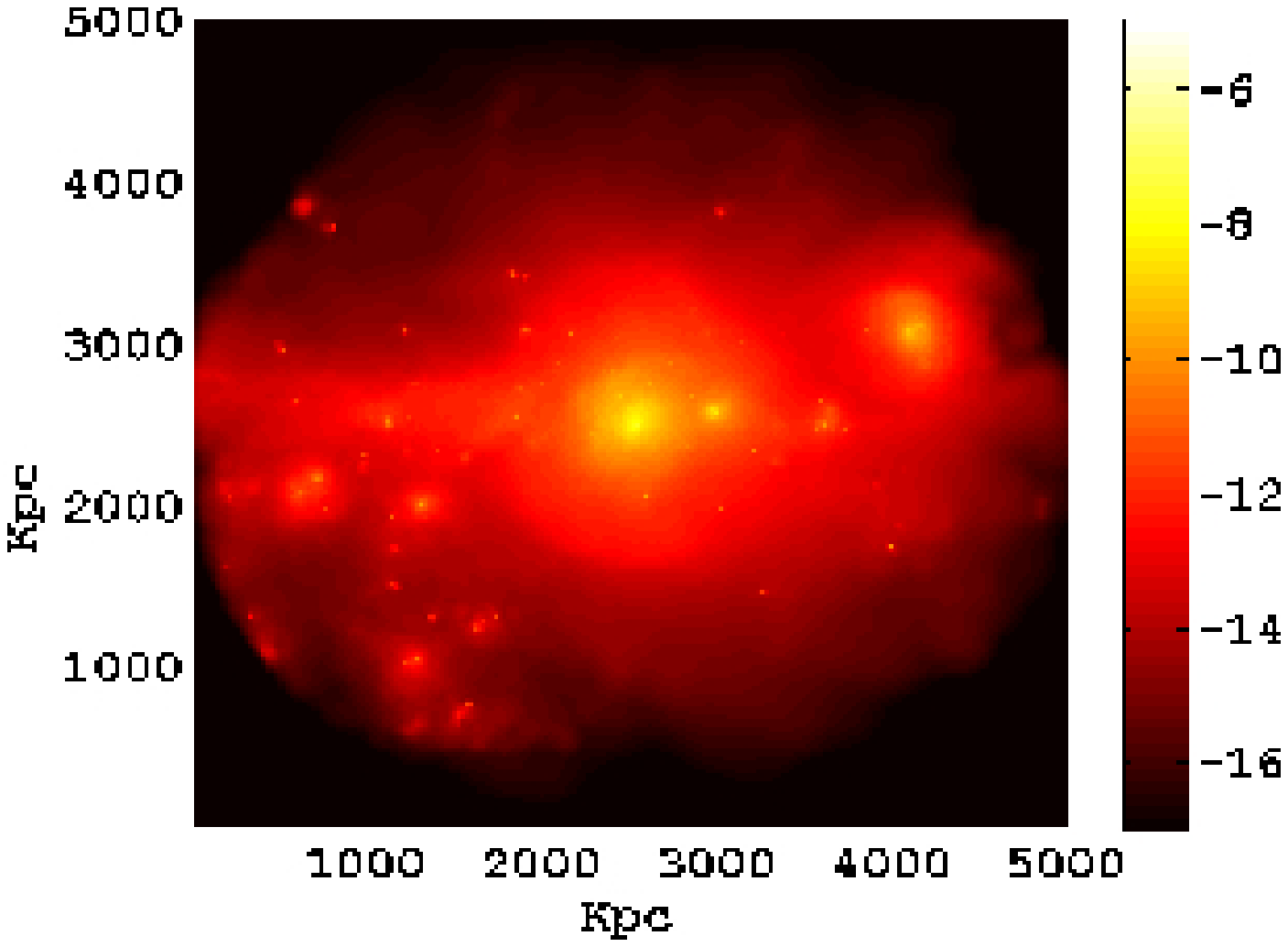}}\\
\resizebox{55mm}{!}{\includegraphics{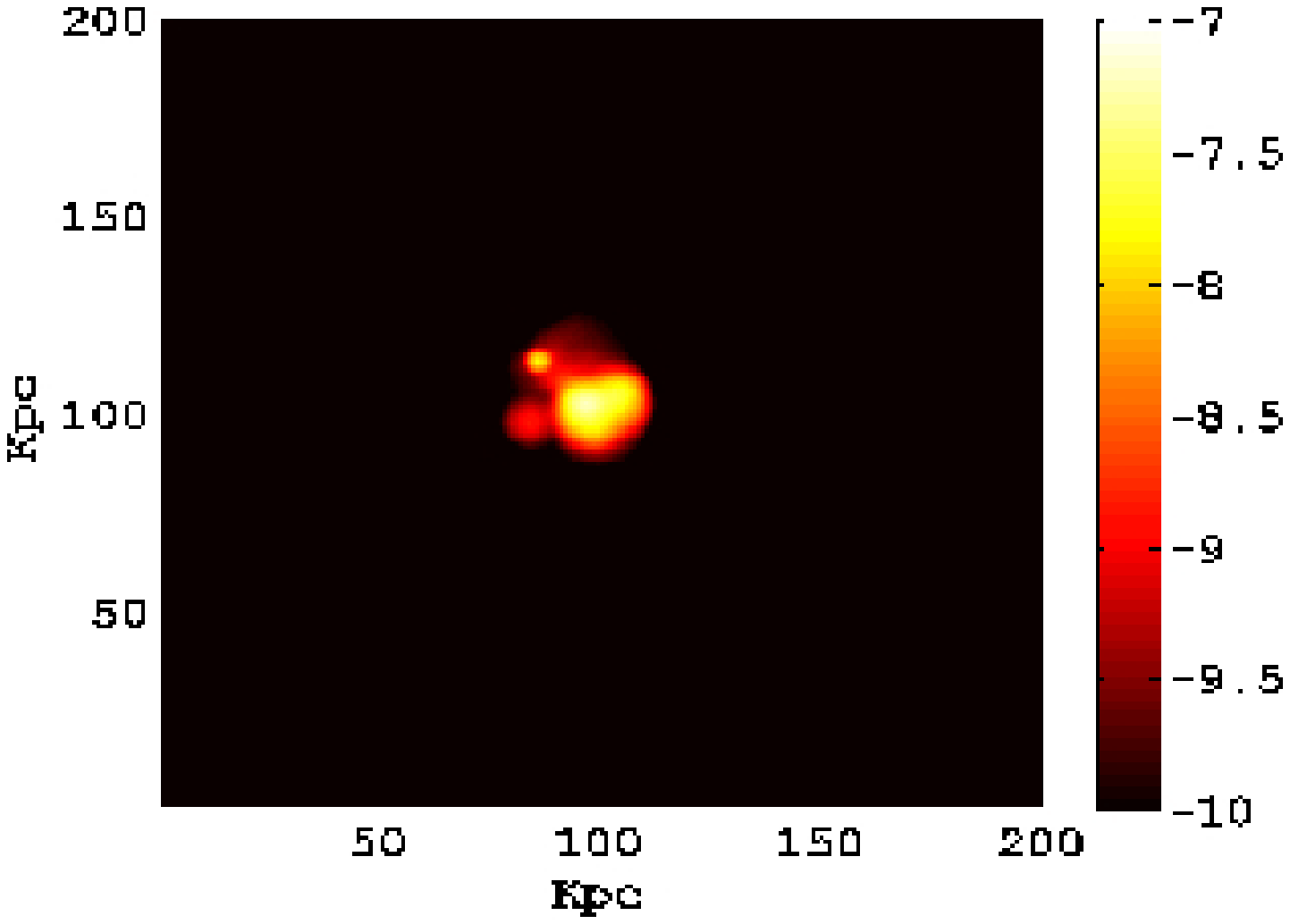}} 
       \resizebox{55mm}{!}{\includegraphics{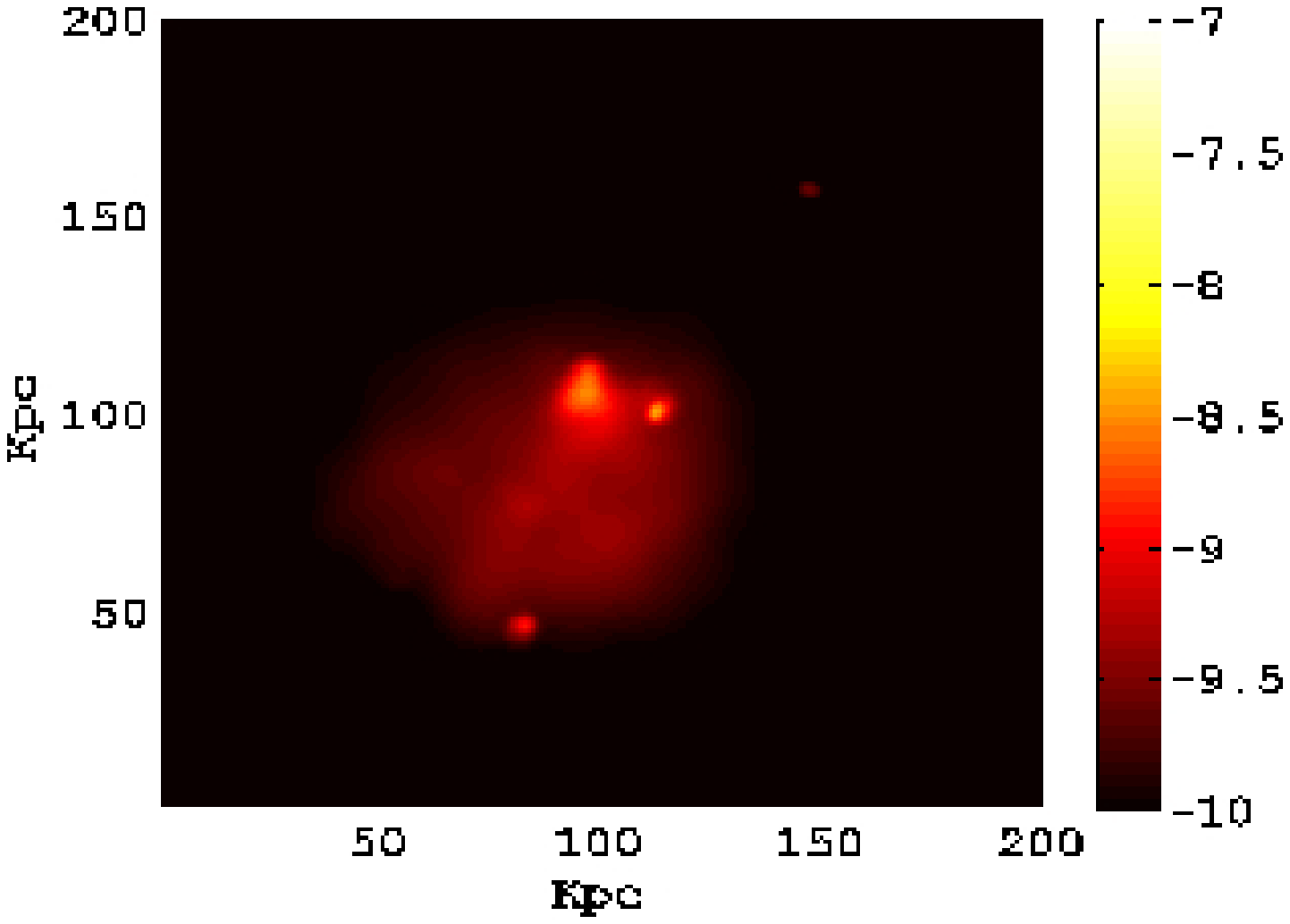}}
        \resizebox{55mm}{!}{\includegraphics{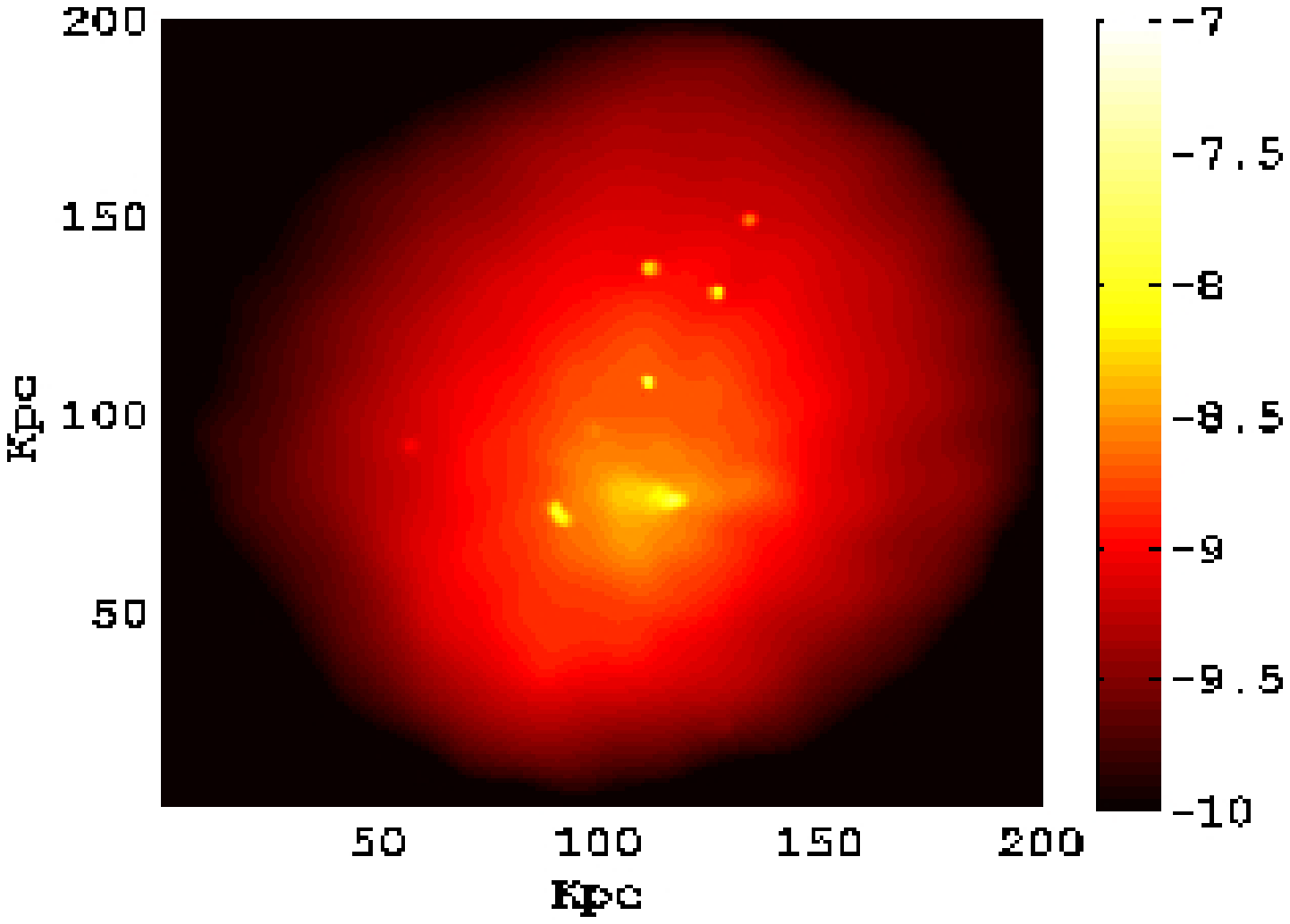}}\\

    \end{tabular}
    \caption{Simulated $y$-distortion maps around two massive black holes,
    at three different redshifts; left, middle, and right columns are for $z=3$, $z=2$, and $z=1$.
     The top two rows are for the most massive black hole in the
    simulation, with mass  $7.35 \times 10^{8} M_{\odot}$, $2.76 \times 10^{9} M_{\odot}$ and $4.26 \times 10^{9}M_{\odot}$ at redshifts 3, 2, and 1 respectively. The top row shows $y$ in a
    5 Mpc square region centered on the black hole; the second row shows zooms in to a smaller
    region 200 kpc square. Note the two rows have different color scales. The third and fourth rows 
    are the same as the first two rows, for a different black hole (the second most massive black hole at redshift 3.0) with mass $7.15 \times 10^{8} M_{\odot}$, $8.2 \times 10^{8} M_{\odot}$ and $2.11 \times 10^{9}M_{\odot}$ at redshifts 3, 2, and 1. For both
    black holes, the peak value of $y$ is between $10^{-7}$ and $10^{-6}$, corresponding to
    an effective maximum temperature distortion between a few tenths of a $\mu$K to a few $\mu$K.}
   \end{center}
\end{figure*}

Typically, the SZ distortions from quasar feedback result in effective
temperature distortions at the micro-Kelvin level on arcminute angular
scales. Recent advances in millimeter-wave detector technology and the
construction of several single-dish and interferometric experiments (see Birkinshaw \& Lancaster 2007 for a recent review on SZ observations), including
the Sub Millimeter Array (SMA), the combined array for research
in Millimeter wave Astronomy (CARMA), the Cornell
Caltech Atacama Telescope (CCAT), the Atacama Large
Millimeter Array (ALMA), and the Large
Millimeter Telescope (LMT), along with SZ surveys like ACT and
SPT, have brought detection of this signal into the realm of
possibility. Although the direct detection of this signal from current SZ
surveys seems unlikely, since the amplitude of fluctuation observed is at or
below the noise threshold of ACT or SPT (Chatterjee \& Kosowsky 2007),
proposed submillimeter facilities offer some possibility for direct detection
of this signal.  An additional route for detection is cross-correlation of
optically-selected quasar with microwave maps (Chatterjee
\& Kosowsky 2007; Scannapieco Thacker \& Couchman 2008). However, this
approach likely requires multifrequency observations to discriminate the SZ
effect from intrinsic quasar emission or infrared sources.
 
The paper is organized as follows. Section 2 describes the simulation that has
been used in this work. In Section 3 we give a brief review of the SZ
distortion and display the SZ maps derived from the simulation. Astrophysical
results from the maps are presented in Section 4, including radial profiles
around individual black holes and the correlation between black hole mass and
SZ distortion. The concluding Section estimates detectability of these signals
and summarizes future prospects.  Throughout we use units with $c=k_B=1$.

\begin{table*}
\begin{tabular}[t]{c|c|c|c|c|c}
\hline
\hline
\multicolumn{1}{c}{}&
\multicolumn{1}{c}{Most massive black hole}&
\multicolumn{1}{c|}{}&
\multicolumn{1}{c}{}&
\multicolumn{1}{c}{Second black hole}&
\multicolumn{1}{c|}{}\\
\hline

\multicolumn{1}{c|}{Redshift}&
\multicolumn{1}{c|}{$N_{BH}$}&
\multicolumn{1}{c|}{$dM/dt$, $M_\odot$/yr}&

\multicolumn{1}{c|}{Redshift}&
\multicolumn{1}{c|}{$N_{BH}$}&
\multicolumn{1}{c|}{$dM/dt$, $M_\odot$/yr}

\\
\multicolumn{1}{c|}{}&
\multicolumn{1}{c|}{}&
\multicolumn{1}{c|}{}&
\multicolumn{1}{c|}{}&
\multicolumn{1}{c|}{}&
\multicolumn{1}{c|}{}

\\
\hline
\hline
 3.0 & 0 & 0.034 & 3.0 & 0 & 0.240  \\ 
 2.0 & 3 & 0.003 & 2.0 & 2 & 0.013 \\ 
 1.0 & 4 & 0.013 &  1.0 & 1 & 0.005\\ 
\hline 
\end{tabular}

\caption{The accretion rates and the number of neighboring black holes within a radius of 100 kpc, for the two black holes in Fig.~1.}
\end{table*}

\begin{figure*}
  \begin{center}
    \begin{tabular}{cc}
    
      \resizebox{60mm}{!}{\includegraphics{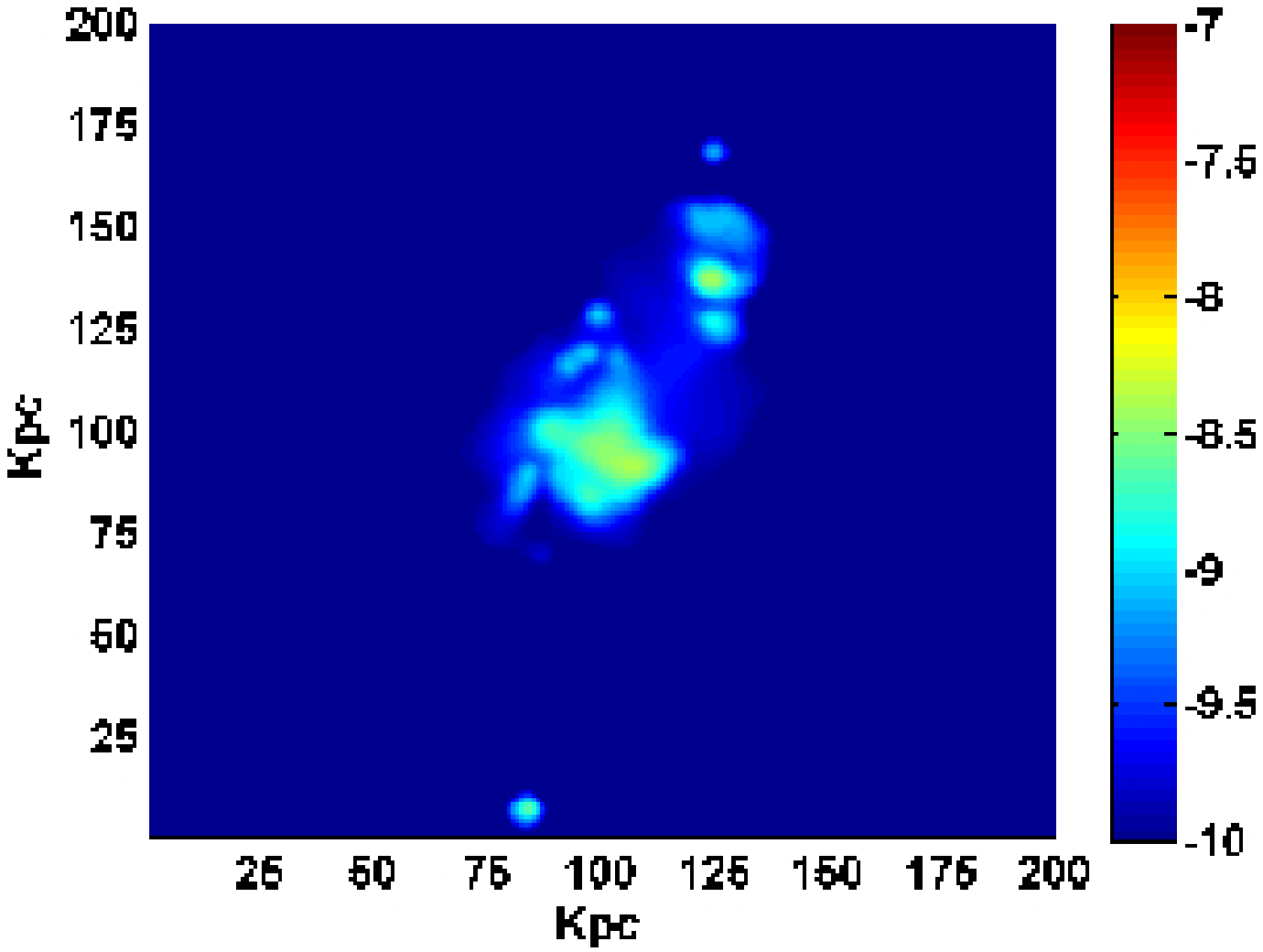}}
       \resizebox{60mm}{!}{\includegraphics{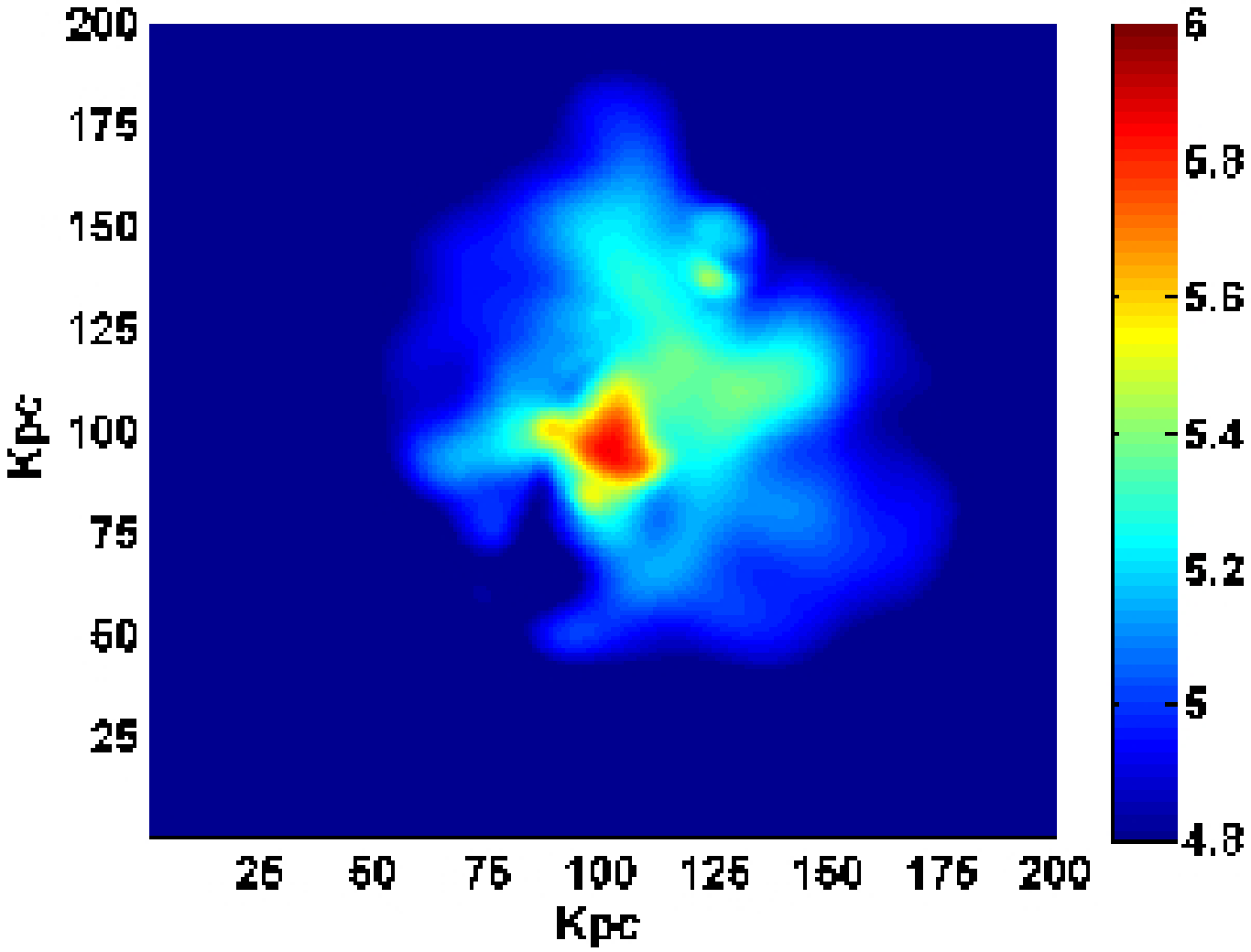}}
       \resizebox{60mm}{!}{\includegraphics{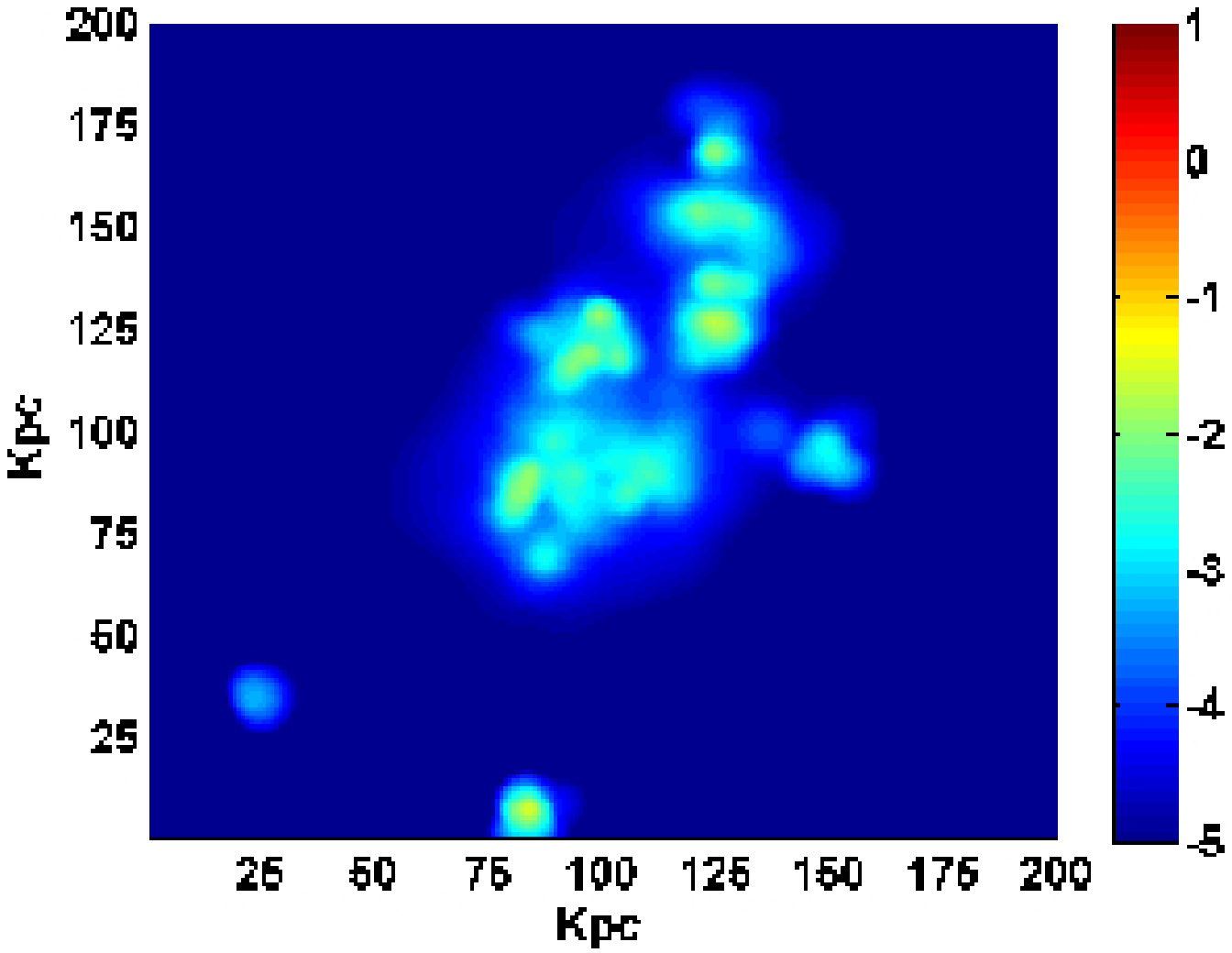}}\\
 \resizebox{60mm}{!}{\includegraphics{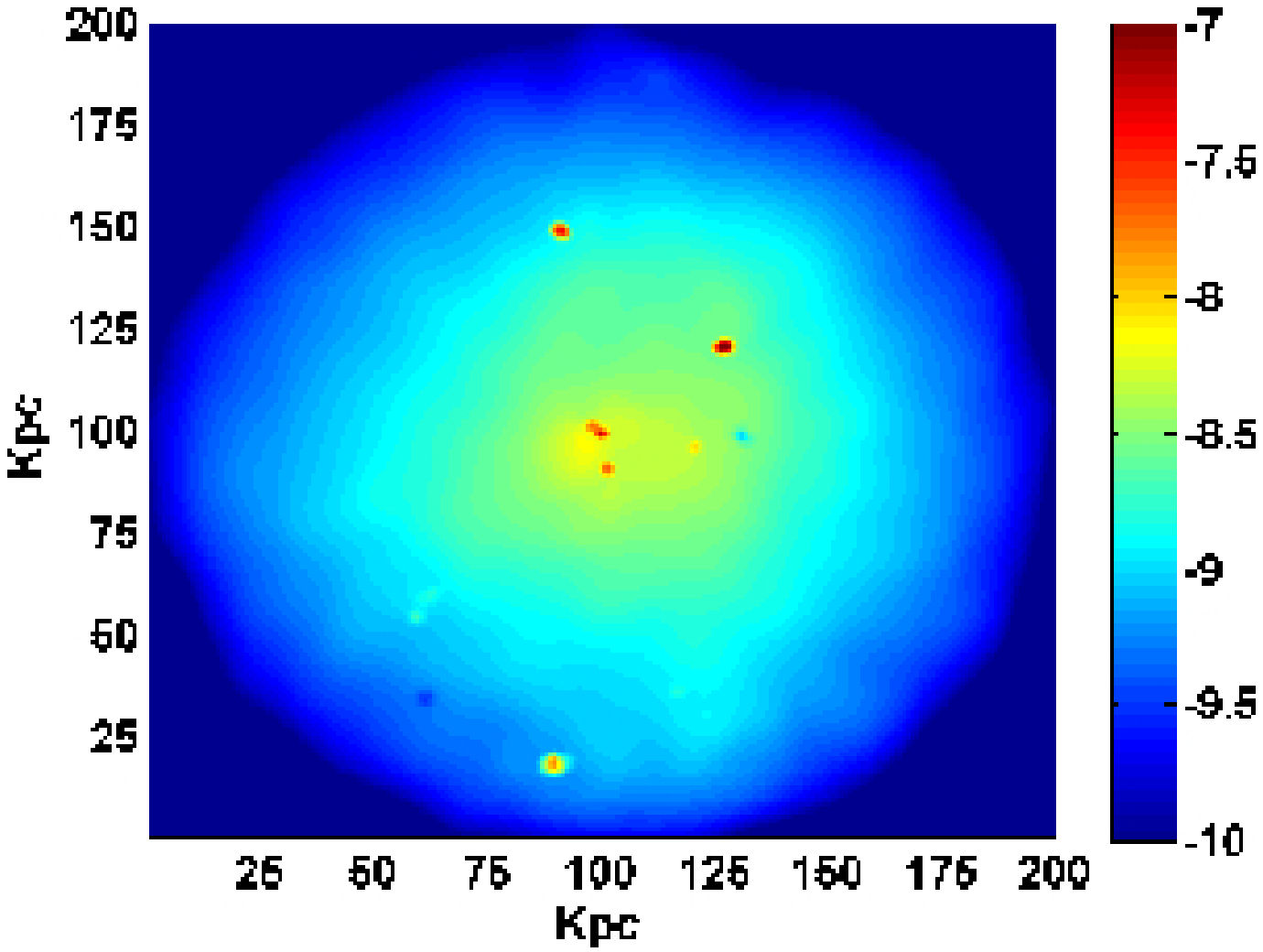}}
       \resizebox{60mm}{!}{\includegraphics{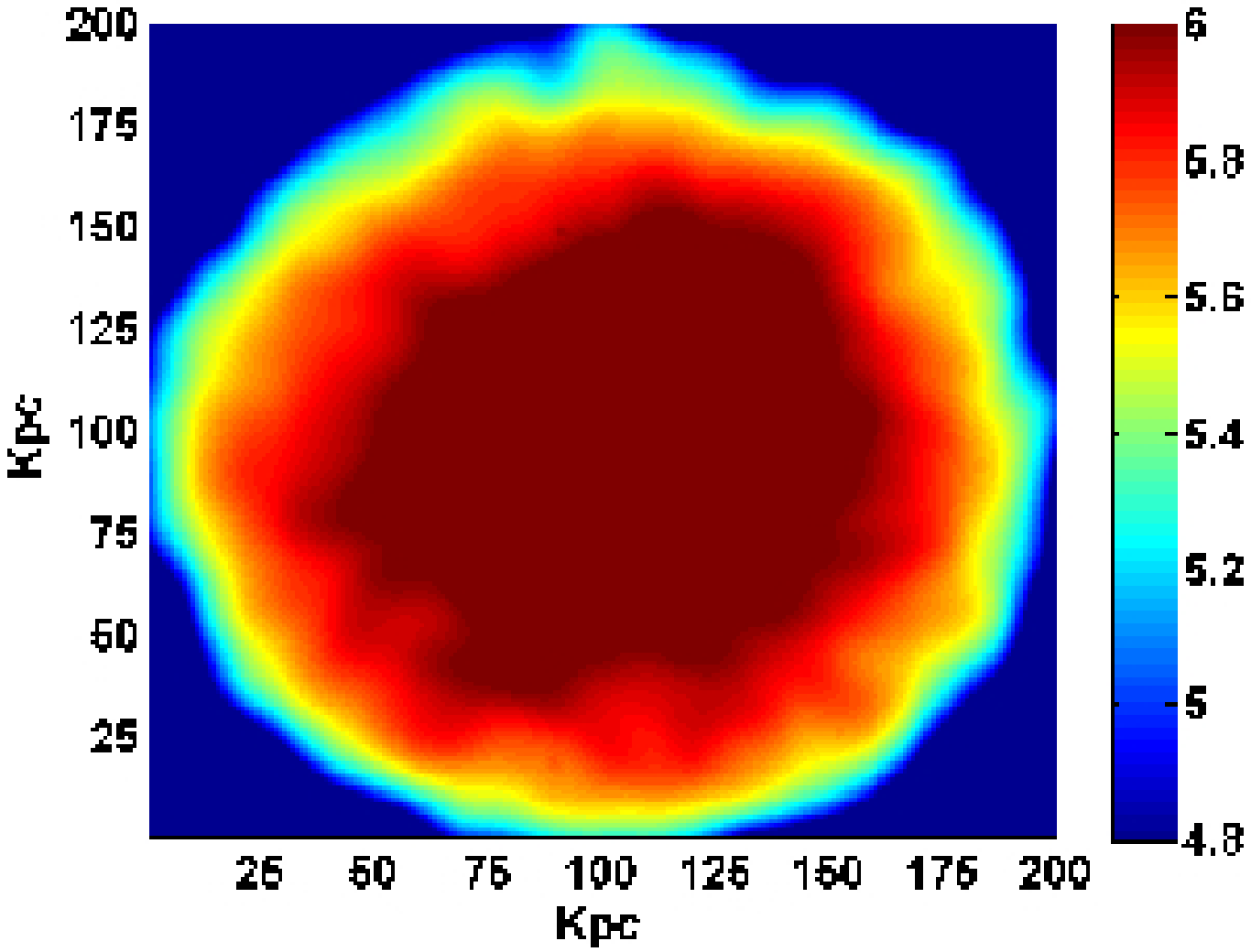}}
       \resizebox{60mm}{!}{\includegraphics{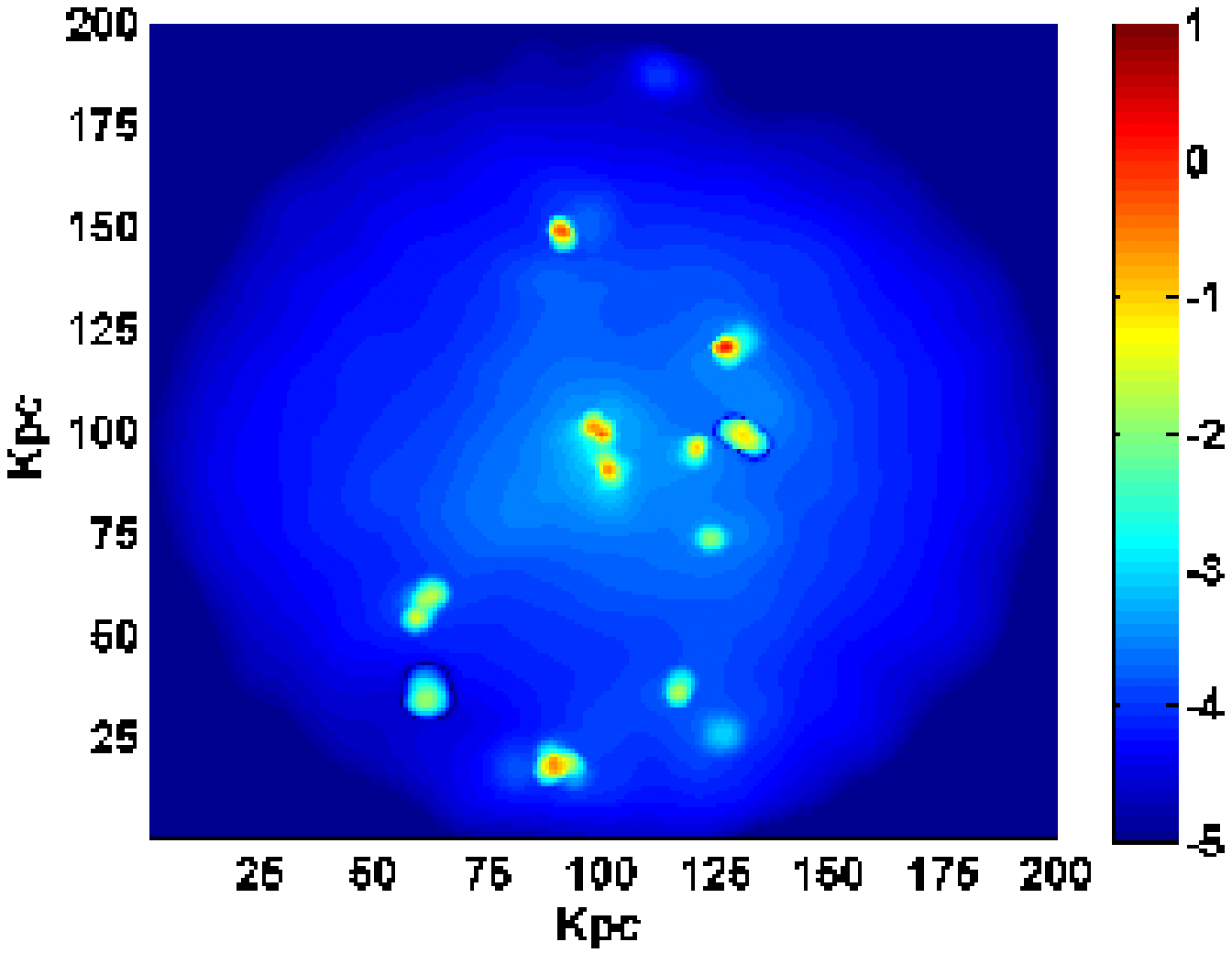}}\\
  \resizebox{60mm}{!}{\includegraphics{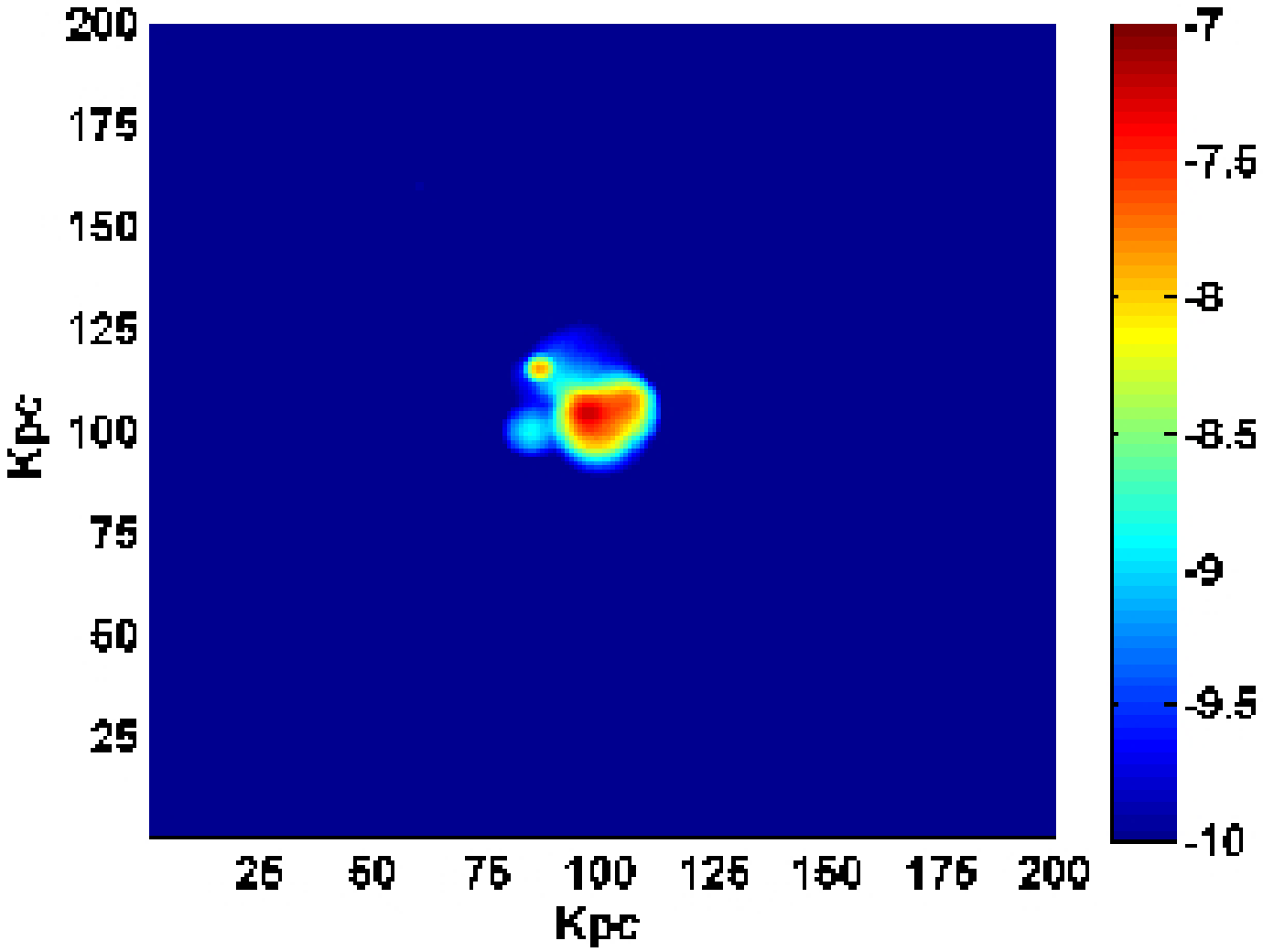}} 
       \resizebox{60mm}{!}{\includegraphics{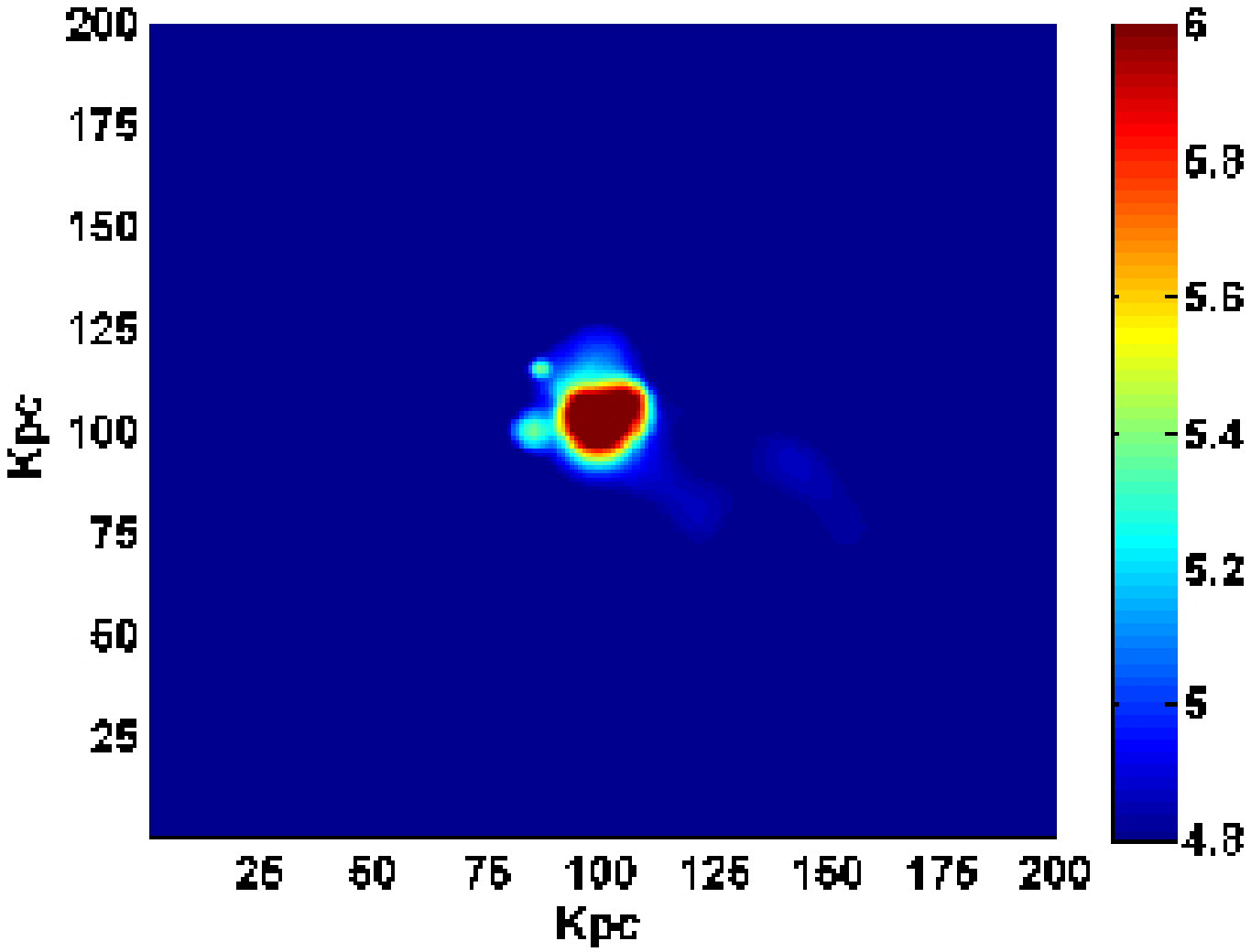}}
        \resizebox{60mm}{!}{\includegraphics{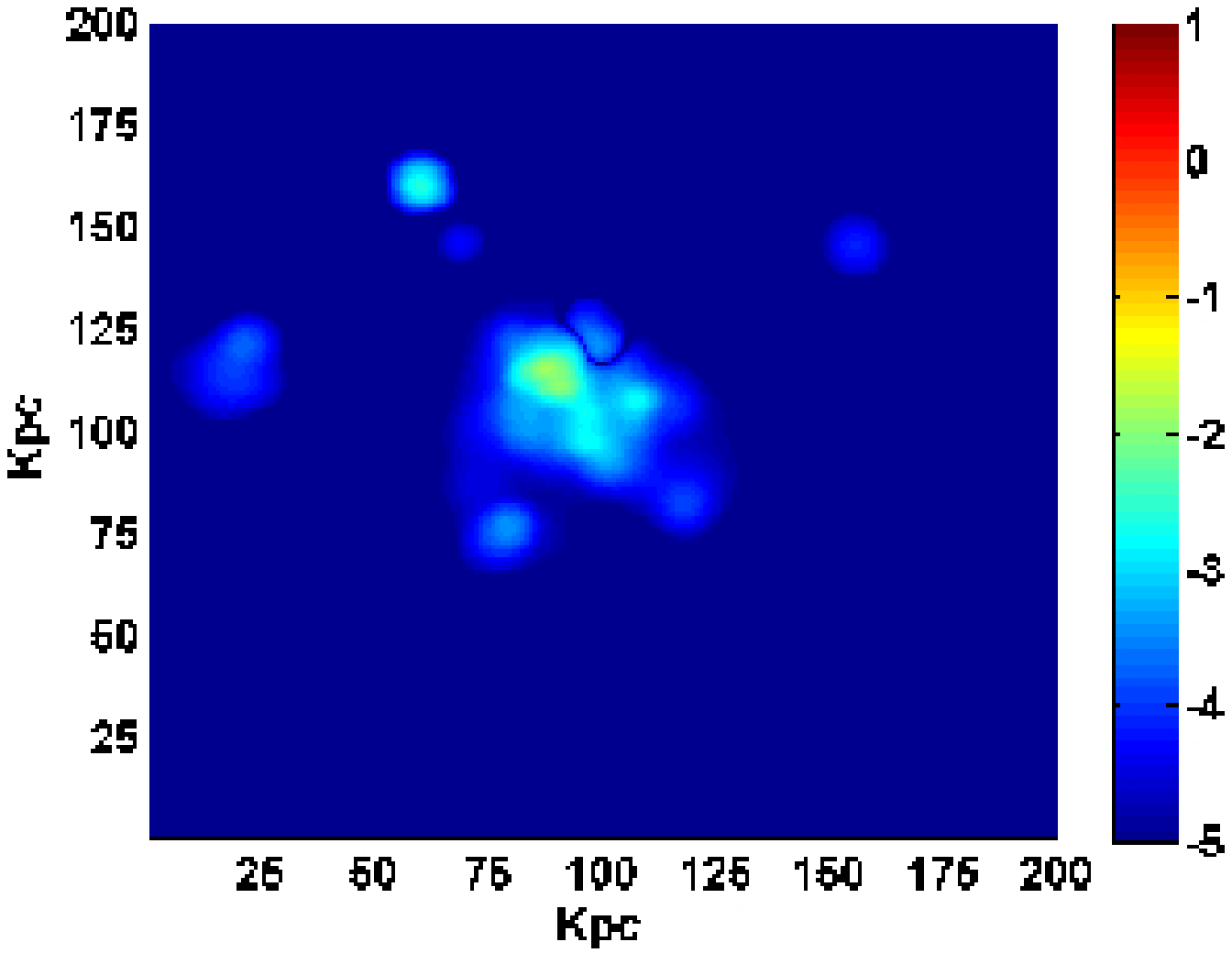}}\\
 \resizebox{60mm}{!}{\includegraphics{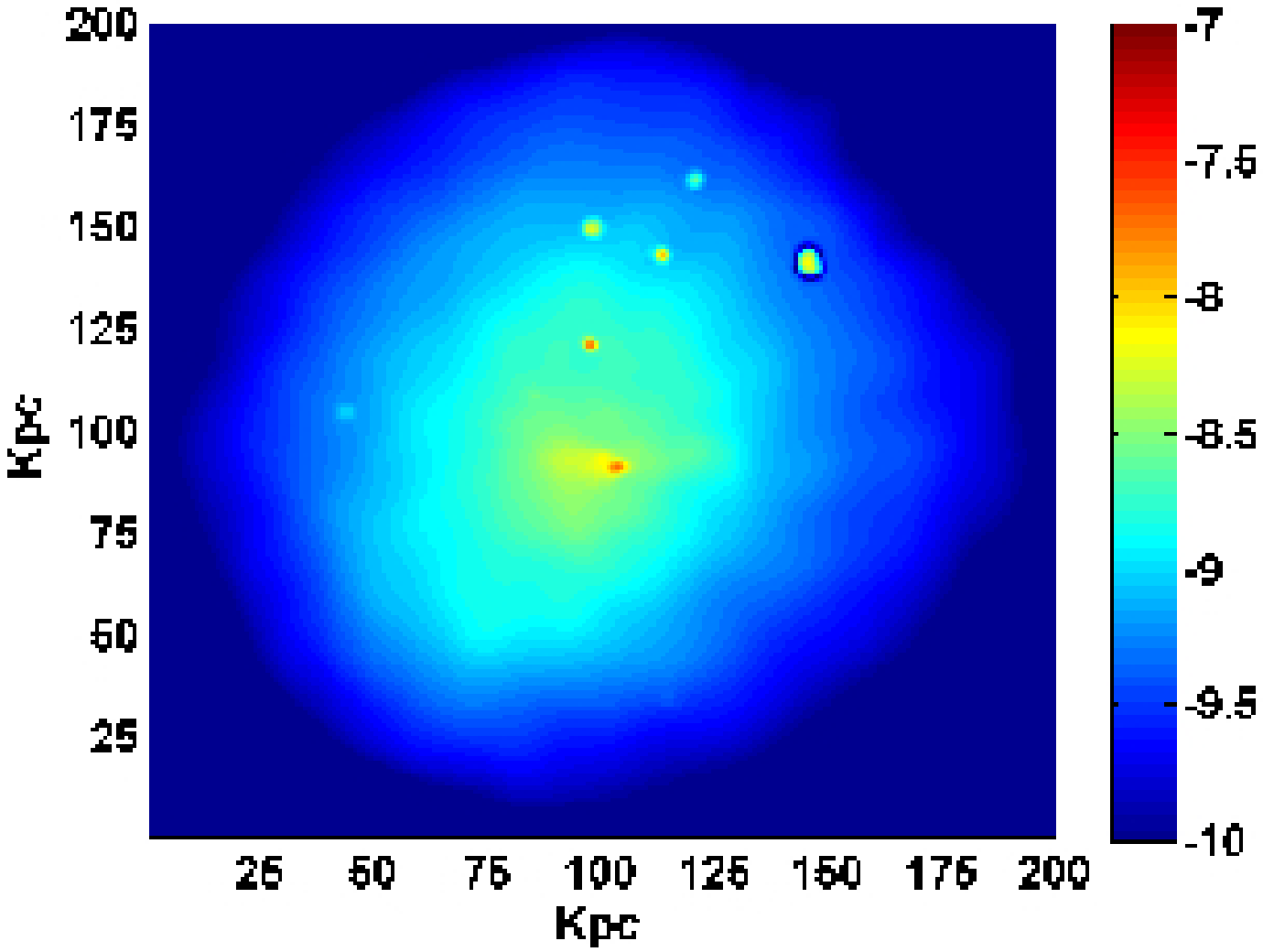}} 
       \resizebox{60mm}{!}{\includegraphics{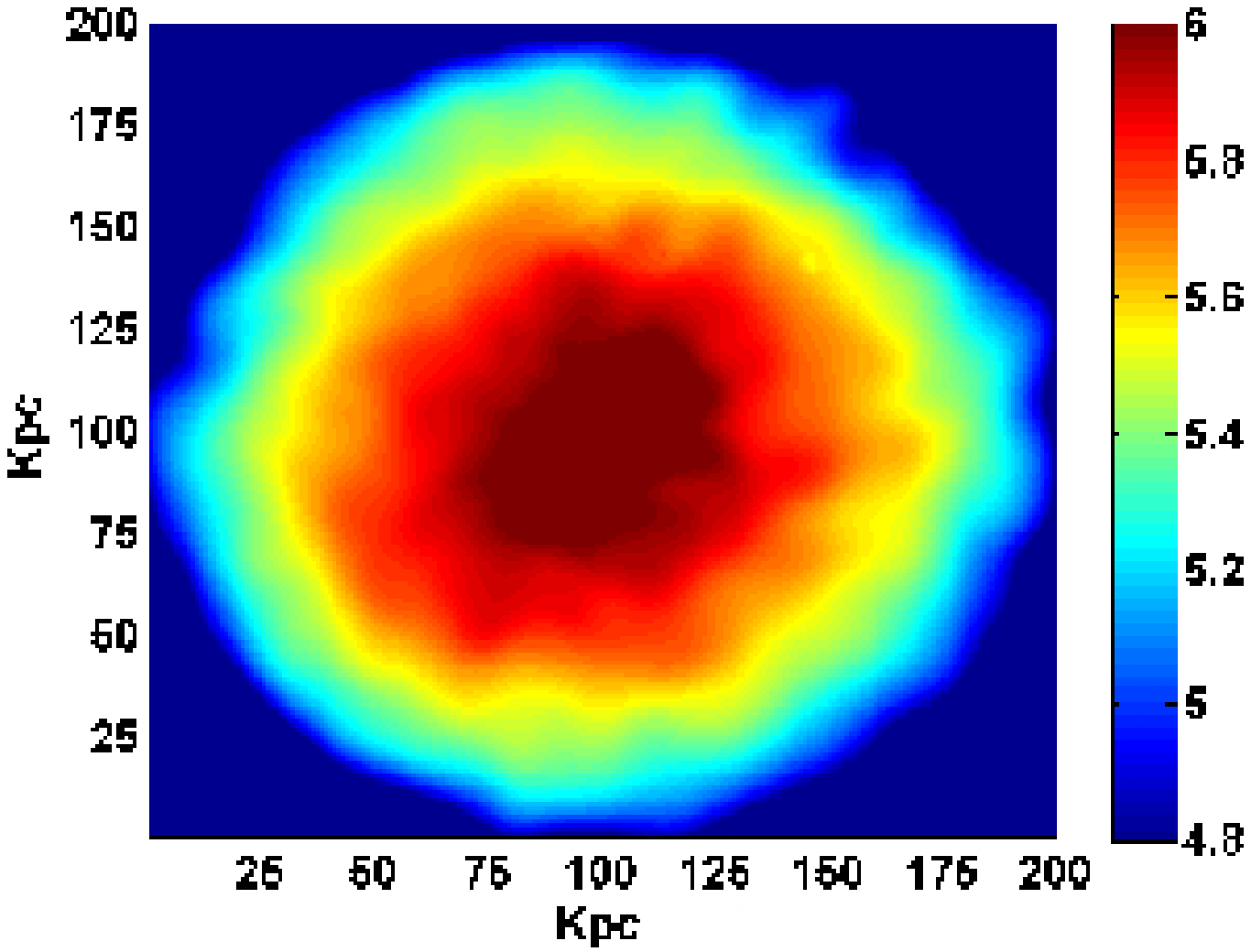}}
        \resizebox{60mm}{!}{\includegraphics{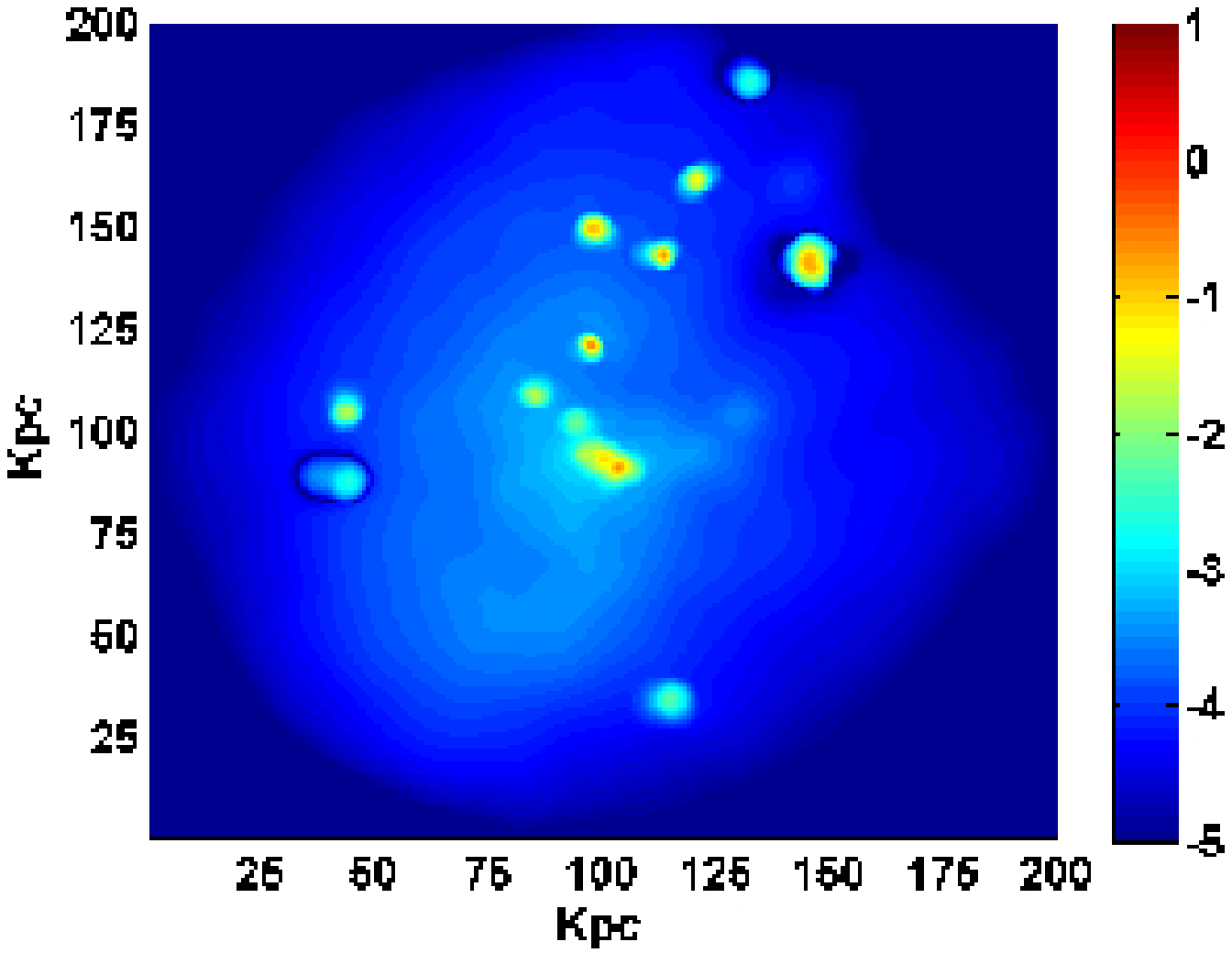}}\\

    \end{tabular}
    \caption{The difference in $y$-distortion between a simulation with black hole feedback and
    a simulation without, for the same region of space shown in Fig.~1.  The two simulations have identical resolution and initial conditions. The first row corresponds to the most massive black hole at $z=3$, the second row at $z=1$; the third row corresponds to the other black hole (second most massive black hole at redshift 3.0) in Fig.~1 at $z=3$, the fourth row at $z=1$. The left column shows $y$, the middle shows the log of the mass-weighted average temperature in units of Kelvin. The right column shows the log of the electron number surface density in units of cm$^{-2}$ .}
   \end{center}
\end{figure*}

\begin{figure*}

    \begin{tabular}{cc}
      \resizebox{90mm}{!}{\includegraphics{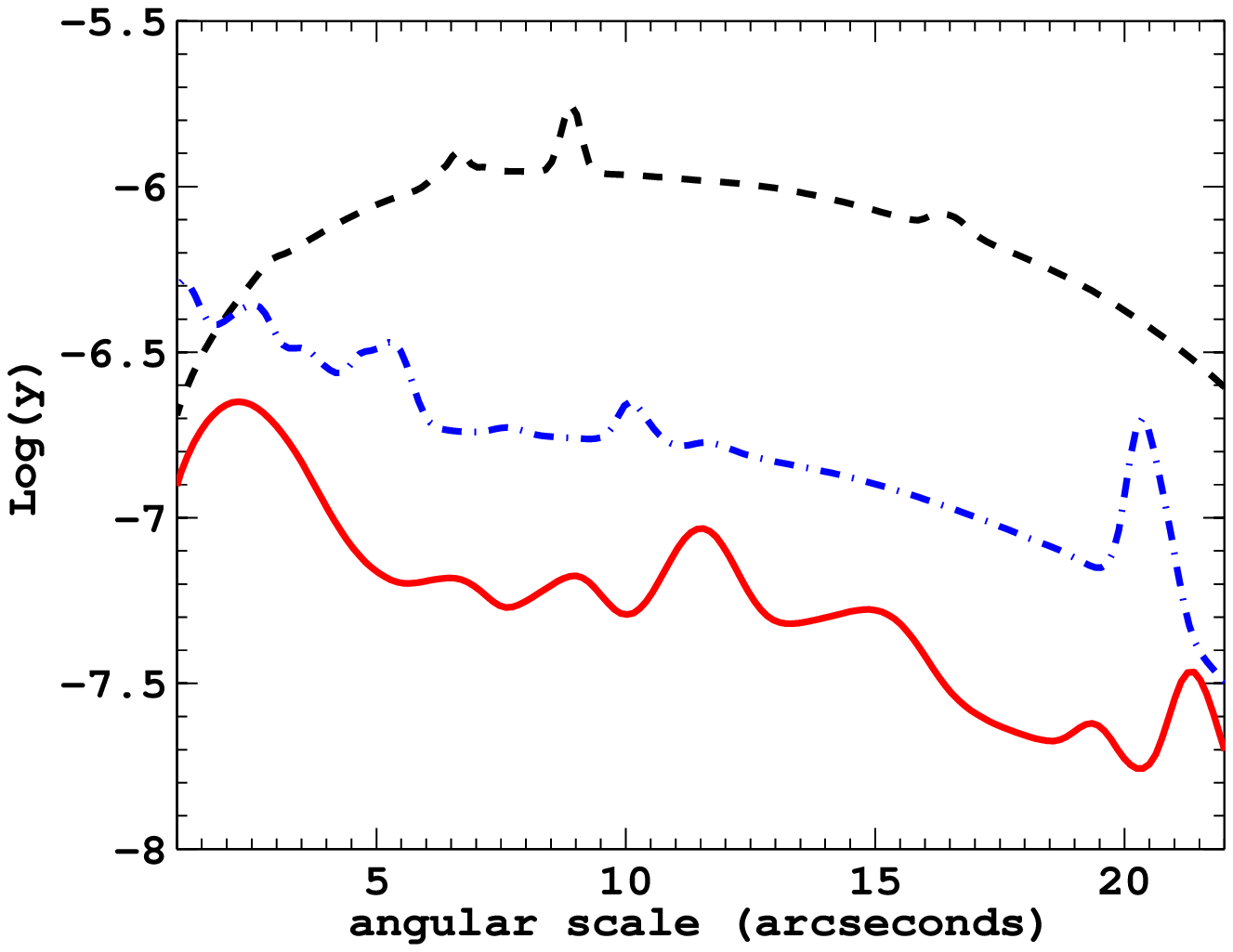}}
      \resizebox{90mm}{!}{\includegraphics{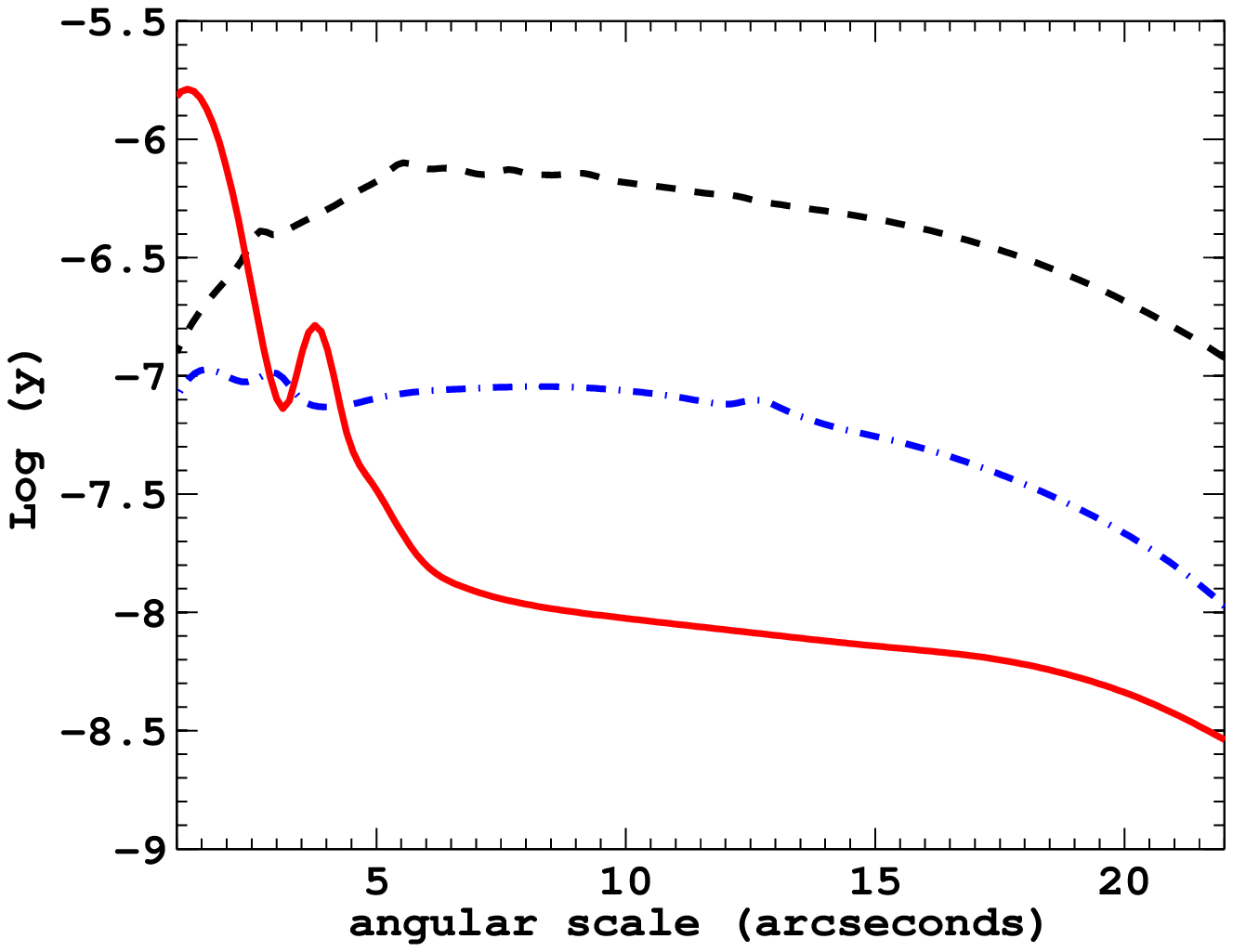}}\\
      
    \end{tabular}
   \caption{The angular profiles of the $y$ distortion for the two black holes shown in Fig.~1 at three different redshifts. The solid, dot-dashed and dashed lines are for redshifts 3, 2, and 1 respectively. The left panel shows the most massive black hole and the right panel is the other black hole (second most massive black hole at redshift 3.0).  Although the central amplitude at a particular time depends strongly on the instantaneous state of the black hole accretion, the distortion amplitude increases monotonically with time at a 20 arcsecond angular distance. }
    \end{figure*}

\section{Simulation}

The numerical code uses a standard $\Lambda$CDM cosmological model with
cosmological parameters from the first year WMAP results (Spergel et al.\
2003). The cosmological parameters are $\Omega_{m} = 0.3$, $\Omega_{\Lambda} =
0.7$, $H_{0} = 70$ km/s Mpc$^{-1}$ and Gaussian initial adiabatic density
perturbations with a spectral index $n_{s}=1$ and normalization
$\sigma_{8}=0.9$. (While the current lower value of $\sigma_8$ will affect the
total number of black holes in a given volume, it should have little impact on
the results for individual black holes presented here.) The simulation uses an
extended version of the parallel cosmological Tree Particle Mesh-Smoothed
Particle Hydrodynamics code GADGET2 (Springel 2005). Gas dynamics are modeled
with Lagrangian smoothed particle hydrodynamics (SPH) (Monaghan 1992);
radiative cooling and heating processes are computed from the prescription
given by Katz, Weinberg, \& Hernquist (1996). The relevant physics of star
formation and the associated supernova feedback has been approximated based on
a sub-resolution multiphase model for the interstellar medium developed by
Springel \& Hernquist (2003a).

A detailed description of the implementation of black hole accretion and the
associated feedback model is given in Di Matteo et al.\ 2008.  Black holes are
represented as collisionless ``sink'' particles that can grow in mass by
accreting gas or by merger events. The Bondi-Hoyle relation (Bondi 1952; Bondi
\& Hoyle 1944; Hoyle \& Lyttleton 1939) is used to model the accretion rate of
gas onto a black hole. The accretion rate is given by $\dot{M}_{BH} =
4\pi[G^{2}M_{BH}^{2}\rho]/(c_{s}^{2} + v^{2})^{3/2}$, where $\rho$ and
$c_{s}$ are density and speed of sound of the local gas, $v$ is the velocity
of the black hole with respect to the gas, and $G$ is the gravitational
constant. The radiated luminosity is taken to be $L_{r} = \eta
(\dot{M}_{BH}c^{2})$ where $\eta=0.1$ is the canonical efficiency for thin
disk accretion.  It is assumed that a small fraction of the radiated
luminosity couples to the surrounding gas as feedback energy $E_{f}$, such
that \(\dot{E_{f}} = \epsilon_{f} L_{T}\) with the feedback efficiency
$\epsilon_f$ taken to be $5\%$.  This feedback energy is put directly into the
gas smoothing kernel at the position of the black hole (Di Matteo et. al
2008). The efficiency $\epsilon_{f}$ is the only free parameter in our quasar
feedback model, and is chosen to reproduce the observed normalization of the $
M_{BH} -\sigma$ relation (Di Matteo, Springel \& Hernquist 2005).  This number
is also consistent with the preheating in groups and clusters that is required
to explain their X-ray properties (Scannapieco \& Oh 2004).  The feedback
energy is assumed to be distributed isotropically for the sake of simplicity;
however the response of the gas can be anisotropic.  This model of quasar
feedback as isotropic thermal coupling to the surrounding gas is likely a good
approximation to any physical feedback mechanism which leads to a shock front
which isotropizes and becomes well mixed over physical scales smaller than
those relevant to our simulations and on timescales smaller than the dynamical
time of the galaxies (see Di Matteo et al.\ 2008 and Hopkins \& Hernquist 2006
for more detailed discussions). In actual active galaxies,
the accretion energy is often released anisotropically through jets. As radio galaxy
lobes can have substantial separations, it is conceivable that actual hot gas bubble morphology
could differ somewhat from that in the simulations. This difference needs to be investigated
with further simulations, but the overall detectability of the signal depends primarily on its
amplitude and characteristic angular scale, which are determined mainly by the total energy
injection as a function of time. The results for the signals and detectability presented
here are unlikely to differ significantly due to more detailed modeling of
the energy injection morphology.

\begin{figure*}

    \begin{tabular}{cc}
      \resizebox{90mm}{!}{\includegraphics{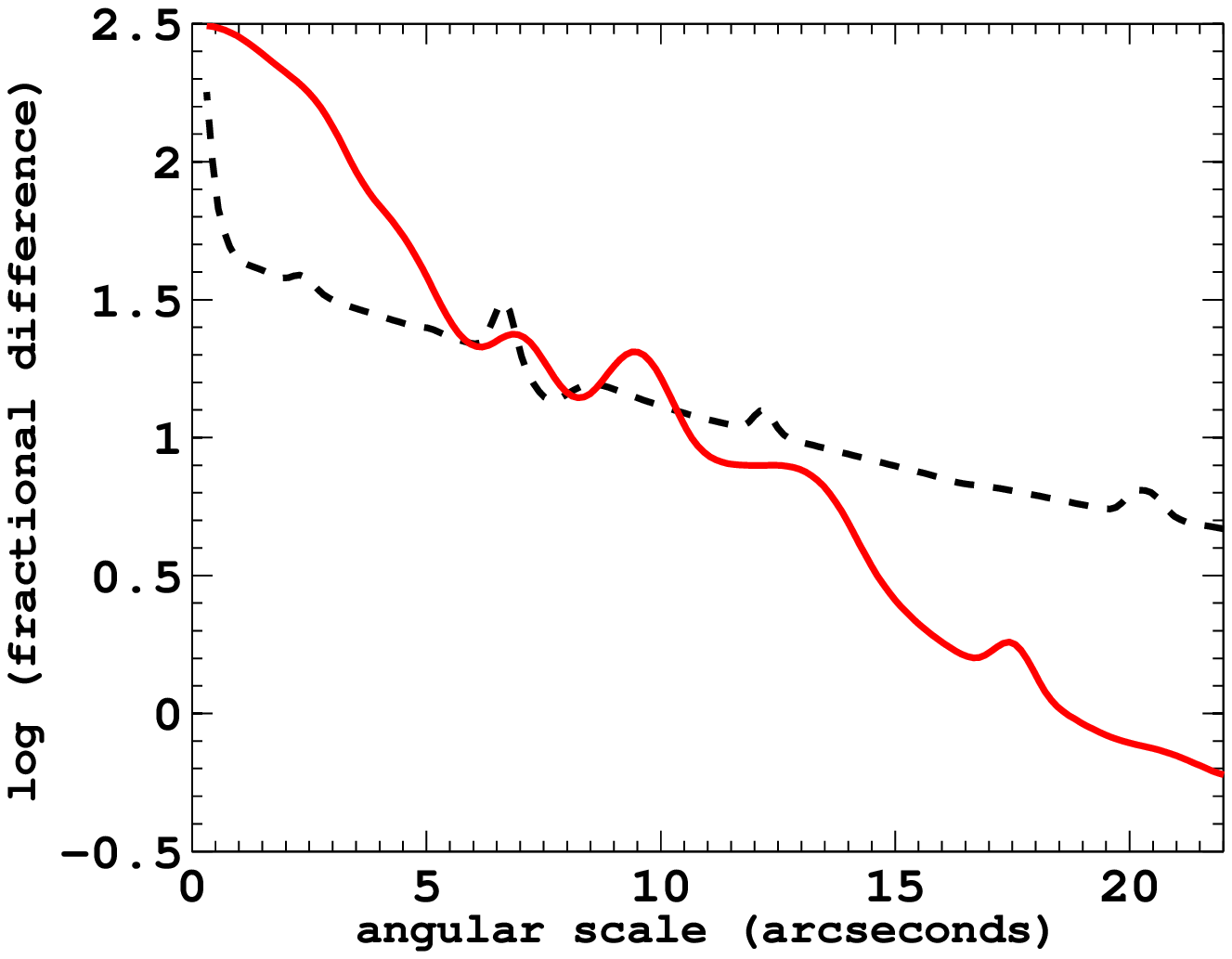}}
      \resizebox{90mm}{!}{\includegraphics{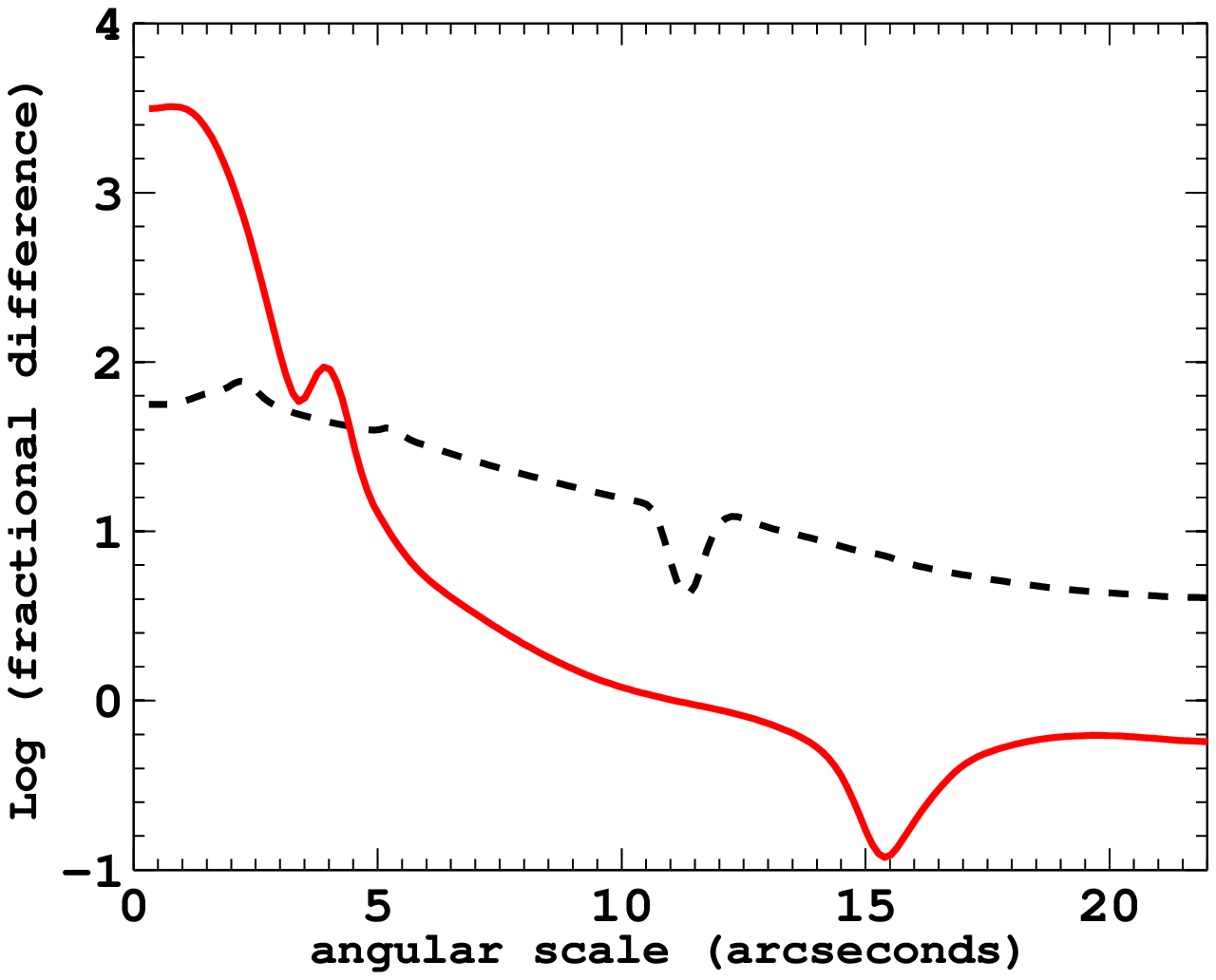}}\\
      
    \end{tabular}
   \caption{The difference in the $y$-distortion radial profile with and
   without black hole feedback, for the two black holes shown in Fig.~1. The
   left panel is for the most massive black hole and the right panel is for
   the second black hole (second most massive black hole at redshift 3.0). For each, the dashed line is the fractional
   change in the y distortion with respect to the no black hole case at $z=1$;
   the solid line is at $z=3$.}
   \end{figure*}

The formation mechanism for the seed black holes which evolve into the
observed supermassive black holes today is not known. The simulation creates
seed black holes in haloes which cross a specified mass threshold. At a given
redshift, haloes are defined by a friends-of-friends group finder algorithm
run on the fly. For any halo with mass $M>10^{10}h^{-1}M_\odot$ which does not
contain a black hole, the densest gas particle is converted to a black hole of
mass $M_{BH}=10^5h^{-1}M_\odot$; the black hole then grows via the accretion
prescription given above and by efficient mergers with other black holes (Di
Matteo et al.\ 2008).
The simulations used in this paper have a box size of $33.75h^{-1}$ Mpc with
periodic boundary conditions. The characteristics of the simulation are listed
in Table 1, where $N_{p}$ is the total number of dark matter plus gas
particles in the simulation, $m_{DM}$ and $m_{\rm gas}$ are their respective
masses, $\epsilon$ gives the comoving softening length, and $z_{\rm end}$ is
the final redshift of the run. For redshifts lower than 1, the fundamental
mode in the box becomes nonlinear, so large-scale properties of the simulation
are unreliable after $z=1$. The current results are derived for the D4 run
with $2\times 216^{3}$ particles; we will present brief comparisons with the
higher-resolution D6 (BHCosmo) run to demonstrate that our results are
reasonably independent of resolution.

A different simulation and feedback model have recently been used by
Scannapieco, Thacker \& Couchman (2008) to study the same issues. They
associate the remnant circular velocity within a post merger event with black
hole mass.  The time scale on which the black hole shines at its Eddington
luminosity is assumed to be a fixed fraction of the dynamical time scale of
the system; the time scale and black hole mass scale are used to estimate the
energy output from a black hole. Their feedback energy efficiency into the
intergalactic medium is 5\%, consistent with the assumption in our
simulation. In contrast, our simulation tracks the time-varying feedback from
a given black hole due to changing local gas density as the surrounding
cosmological structure evolves. This simulation offers the possibility of
tracking the accretion history and duty cycle of black hole emission for
individual black holes, which we plan to address in future work.

  \begin{figure*}
    \begin{tabular}{c}
       \resizebox{60mm}{!}{\includegraphics{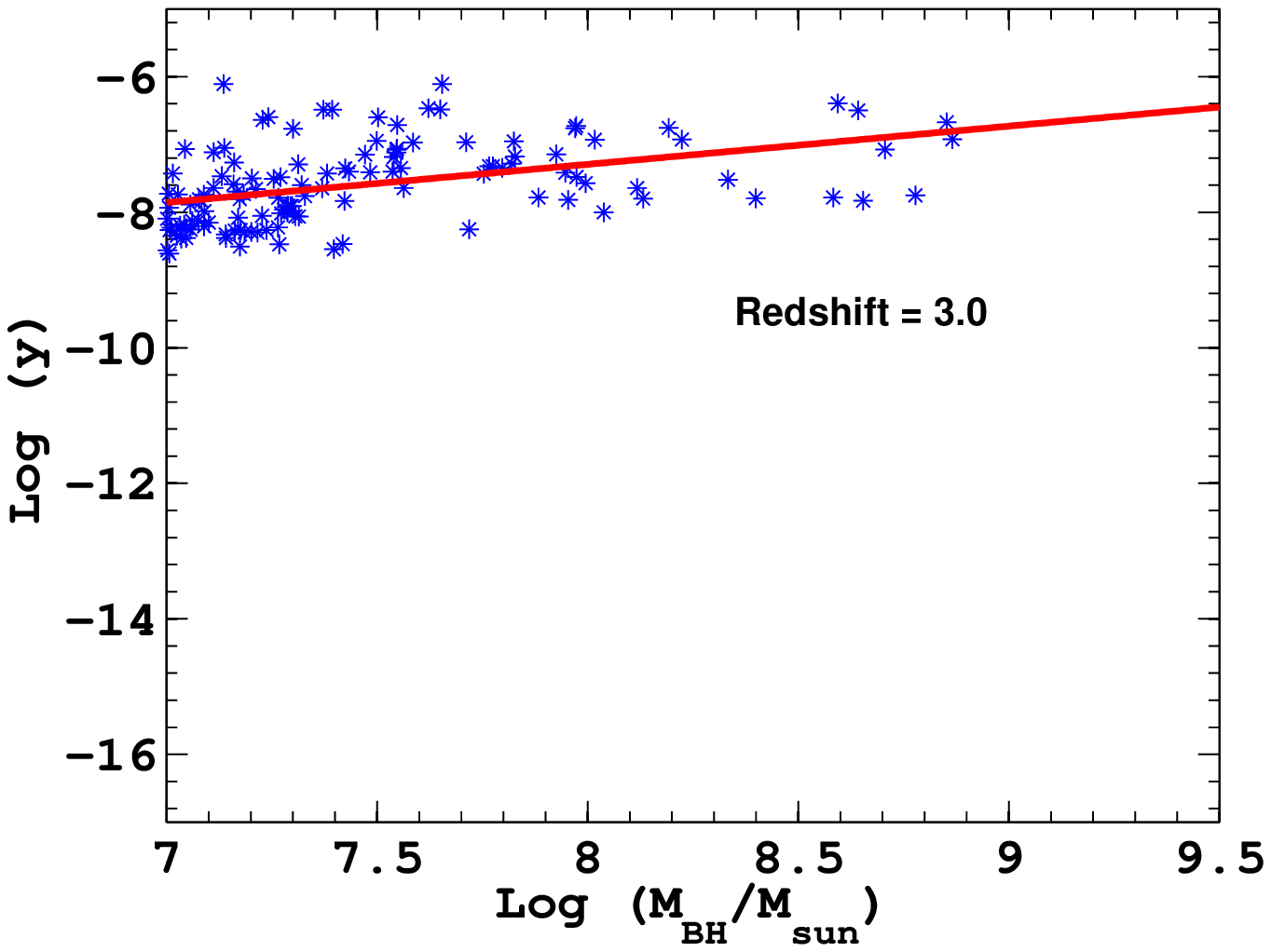}} 
        \resizebox{60mm}{!}{\includegraphics{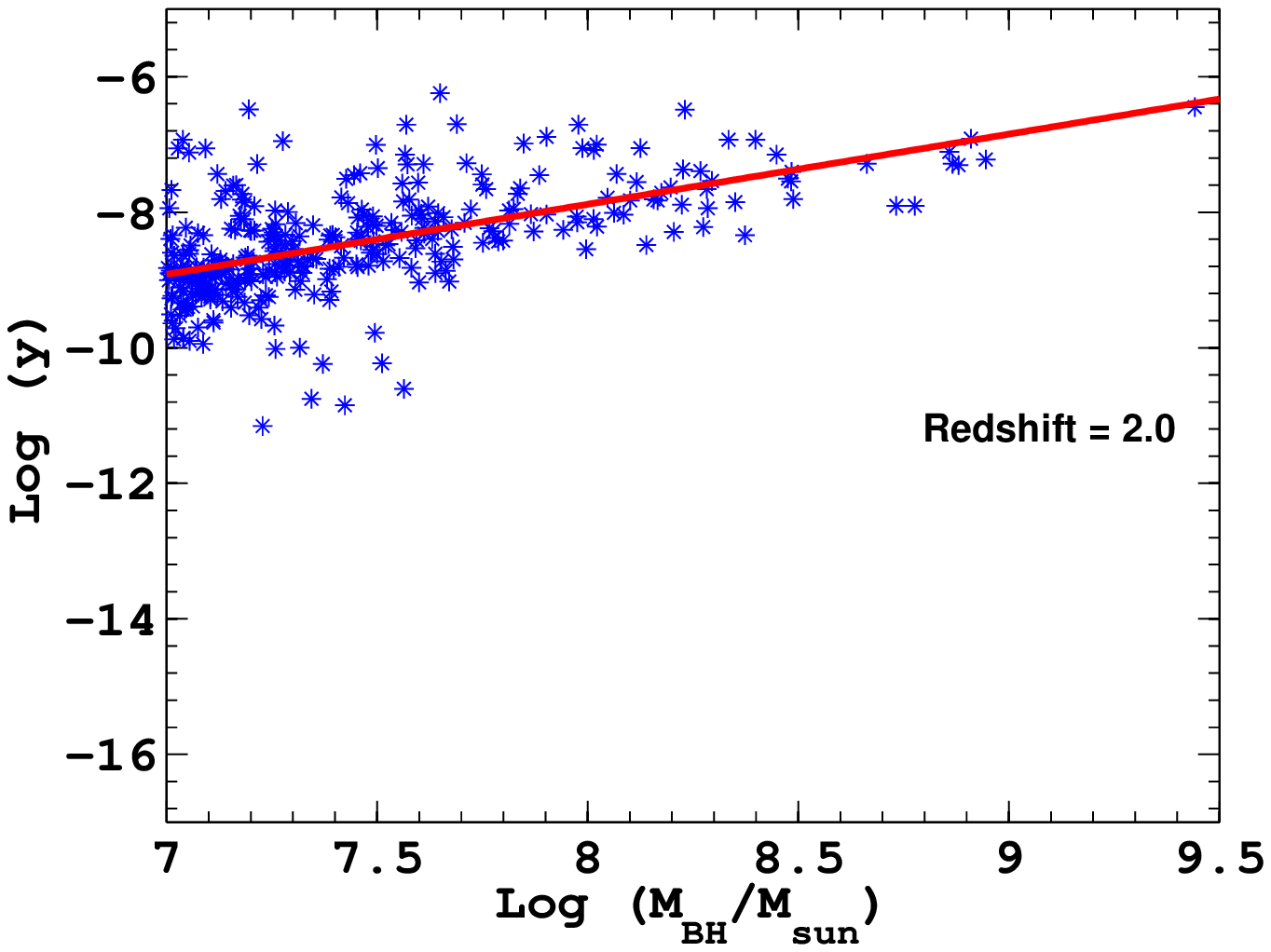}} 
        \resizebox{60mm}{!}{\includegraphics{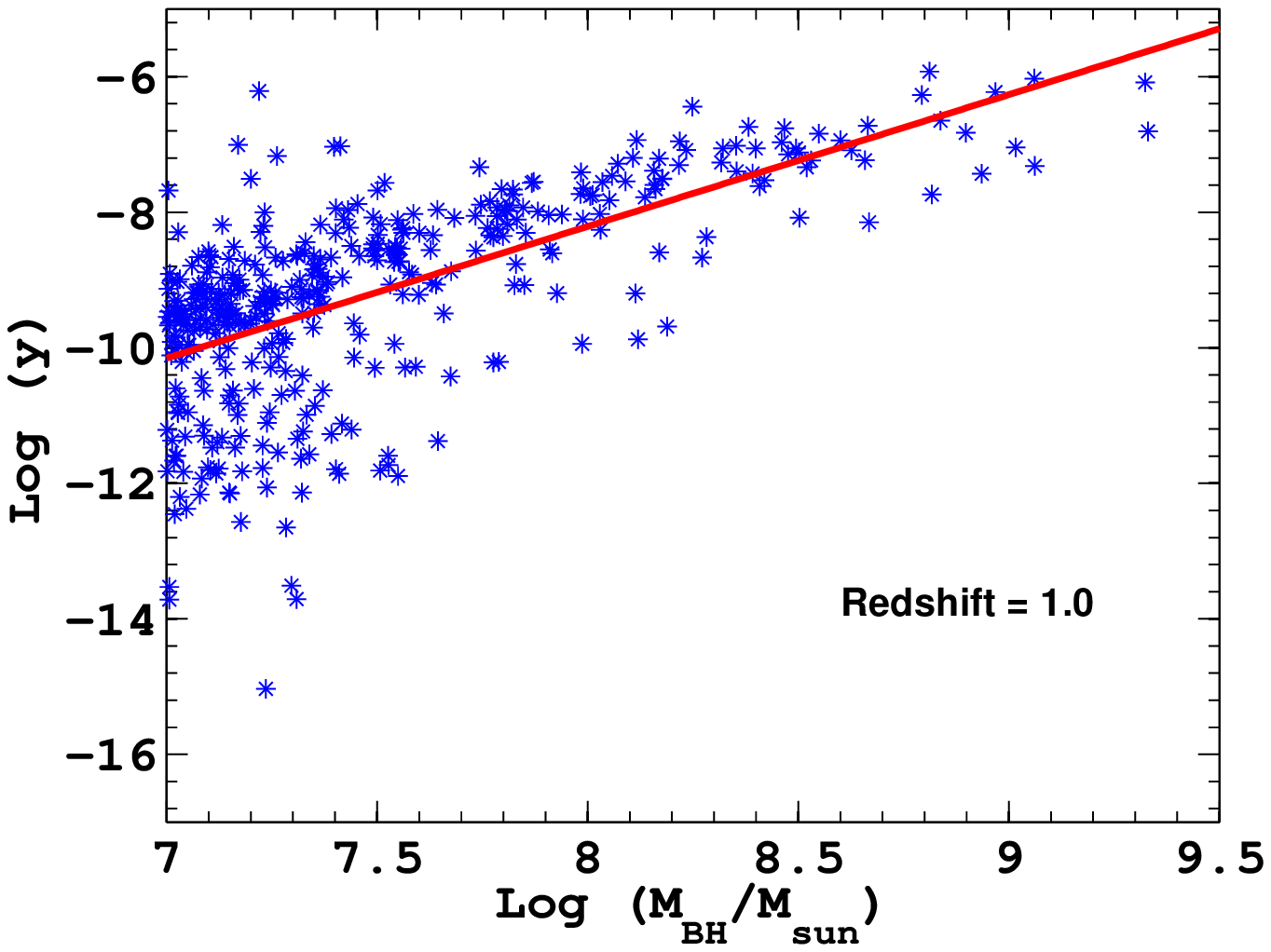}}\\
 \resizebox{60mm}{!}{\includegraphics{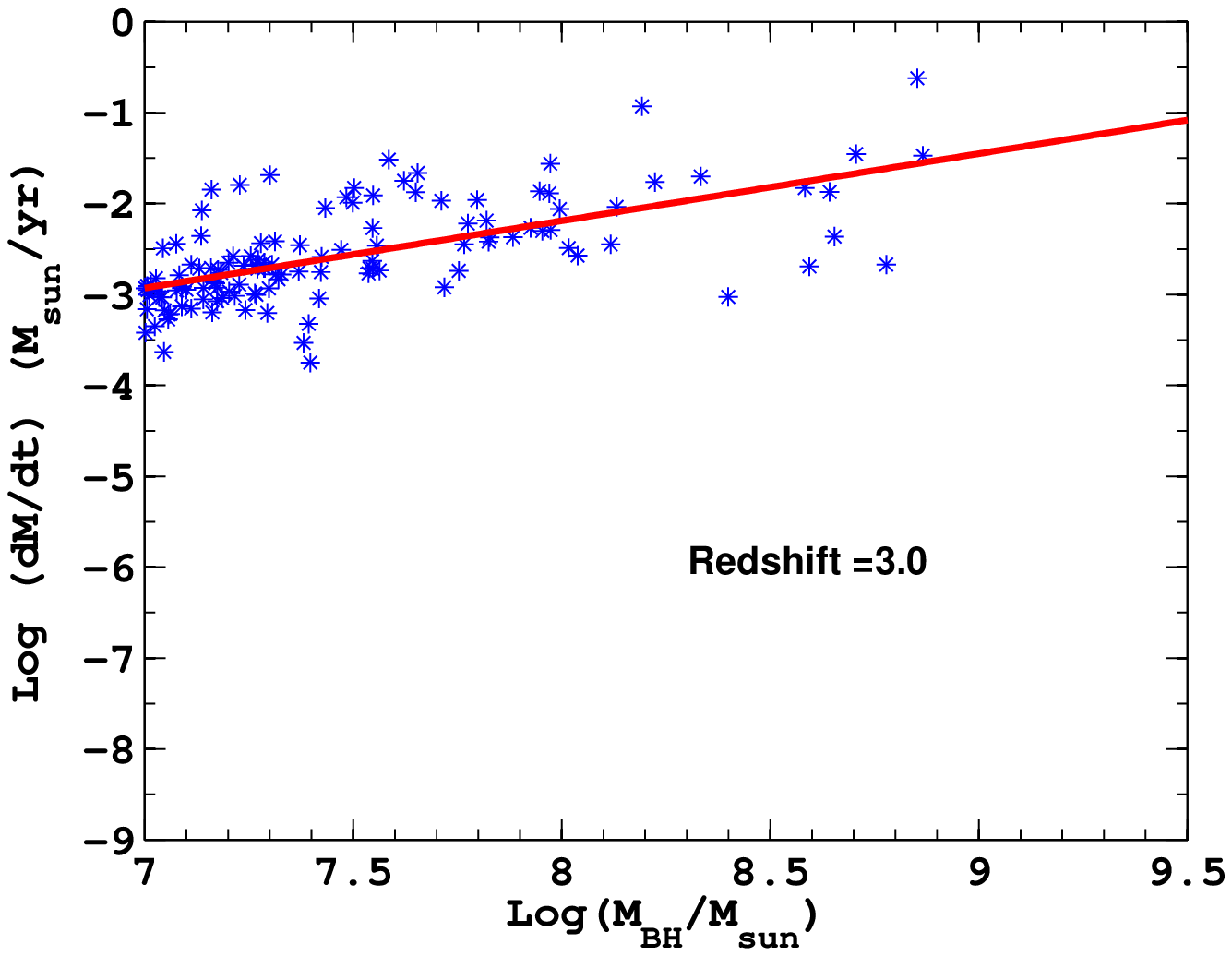}} 
        \resizebox{60mm}{!}{\includegraphics{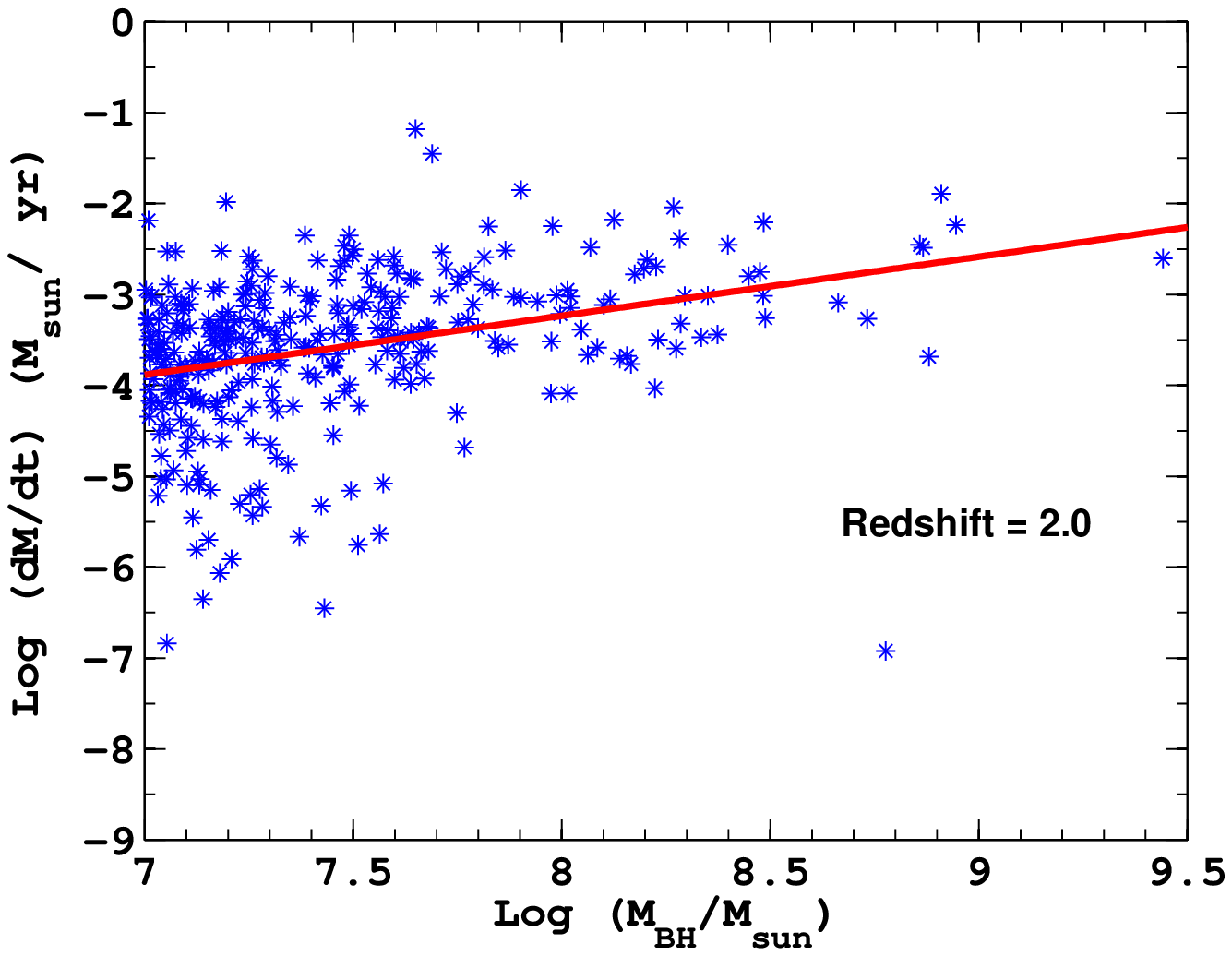}} 
        \resizebox{60mm}{!}{\includegraphics{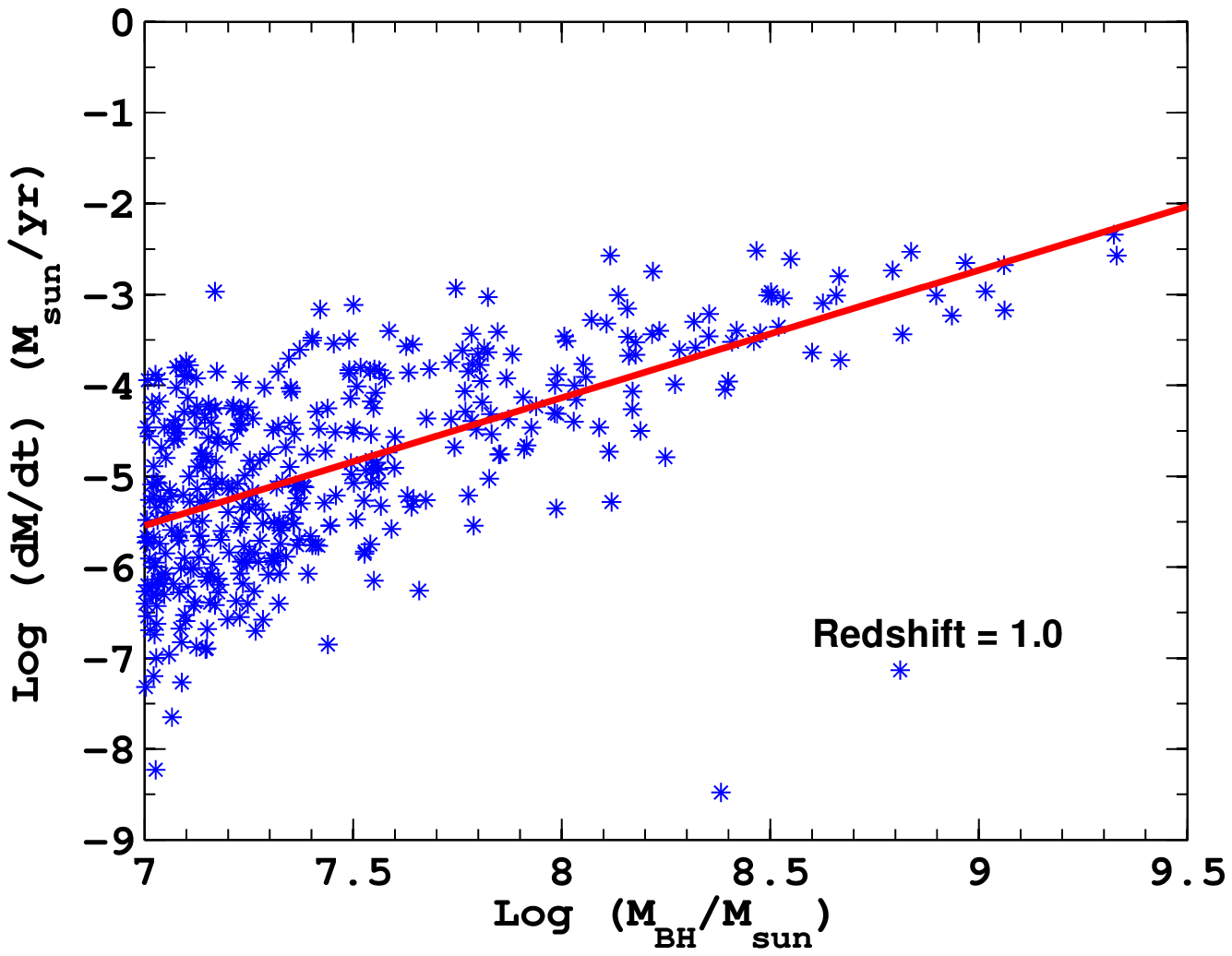}}\\
\end{tabular}
    \caption{The top row shows the mean $y$ distortion within a comoving 200
    kpc region around the black hole as a function of black hole mass, for
    redshifts 3,2,1 from left to right; at $z=1$ the angular size of the box
    is around 25 arcseconds. The bottom row shows the mass accretion rate as a
    function of black hole mass, for the same redshifts. The points are
    the numerical data and the solid lines are power-law fits. All black
    holes in the plotted mass range are included.  The qualitative similarity
    between the top and the bottom panel shows the association of the
    $y$-distortion with accretion rates.  }
\end{figure*} 
\begin{table*}

\begin{tabular}{c|c|c|c|c}
\hline
\multicolumn{1}{c|}{Redshift}&
\multicolumn{1}{c|}{$N_{BH}$}&
\multicolumn{1}{c|}{$N_{BH}$above $10^{7}M_{\odot}$}&
\multicolumn{1}{c|}{Fits for $y$}&
\multicolumn{1}{c|}{Fits for $dM_{BH}/dt$}\\
\hline
 3.0 & 2378 & 127 & $\log y = 0.56 \log(M_{BH}/M_{\odot})-9.8$ & $\log(dM_{BH}/dt) = 0.74 log(M_{BH}/M_{\odot})-8.1$ \\ 
 2.0 & 3110 & 336 & $\log y = 1.00 \log(M_{BH}/M_{\odot})-14$ & $\log(dM_{BH}/dt) = 0.65 logM_{BH}/M_{\odot})-8.4$\\ 
 1.0 & 3404 & 404 & $\log y = 1.90 \log(M_{BH}/M_{\odot})-22$ & $\log(dM_{BH}/dt) = 1.4 log(M_{BH}/M_{\odot})-15$\\ 
 \hline 
\end{tabular}
\caption{Numerical values used in Fig.~5. Column 2 shows the total number of black holes in the simulation at redshifts 3, 2, and 1, while column 3 shows the total number of black holes above a mass of $10^{7}M_{\odot}$. Columns 4 and 5 show the scaling relations displayed in Fig.~5. The mass accretion rate is in units of $M_\odot$/yr. }
\end{table*}


\section{Results from the simulations}
\subsection{The Sunyaev-Zeldovich Distortion and Maps}

The Compton $y$-parameter characterizing the non-relativistic thermal SZ spectral distortion is proportional to
the line-of-sight integral of the electron pressure: 
\begin{equation} 
y = 2\int dl \,\sigma_{T}n_{e}\frac {T_{e}}{m_{e}}
\label{ydef}
\end{equation}
where $\sigma_{T}$ is the Thompson cross section, $n_{e}$ and $T_{e}$ are
electron number density and temperature, and the integral is along the line of
sight.  The effective temperature distortion at a frequency $\nu$ is given by
(Sunyaev \& Zeldovich 1972)
\begin {equation}
\frac{\Delta T}{T_{0}} = \left[x\coth(x/2)-4\right]y,
\end{equation}
where $x=h\nu/T_0$ and $T_{0}$ is the CMB temperature equal to 2.73 K.
  
Figure 1 shows $y$-distortion maps centered around two representative black
holes in the simulation at redshifts 3, 2 and 1 (from left to right
respectively). The two black holes are the most massive and the second most massive black hole at redshift 3.0 in the simulation. We have chosen the two most massive black holes in the simulation since the amplitude of the SZ distortion from the most massive black holes is relevent within the realm of current and future experiments. These maps were made by evaluating the line-of-sight integral
in Eq.~\ref{ydef} through the appropriate portion of the simulation box. In
order to characterize the large scale structure and associated $y$-distortions
surrounding the black holes, we show a large region of the simulation within a
comoving radius of 2.5 Mpc of the black hole in question, displayed with a
comoving box size of 5 Mpc (top and third row for the most massive black hole
and for another black hole in the simulation respectively) as well as a zoom
into the central 200 kpc box (second and forth rows).  The smaller region (200
Kpc) is the relevent scale of interest when looking at the direct impact of
the central black hole to its surrounding gas; in the larger box multiple
black holes are present. The mass of the central black hole is $7.35\times
10^{8}M_{\odot}$ at $z=3$, $2.76\times 10^{9}M_{\odot}$ at $z=2$, and
$4.32\times 10^{9}M_{\odot}$ at $z=1$ (top two rows) and $7.11\times
10^{8}M_{\odot}$ at $z=3$, $8.2\times 10^{8}M_{\odot}$ at $z=2$, and
$2.11\times 10^{9}M_{\odot}$ at $z=1$ (third and forth row).  The feedback
energy associated with black hole accretion creates a hot bubble of gas
surrounding the black hole, which, as shown in the figures, grows
significantly in size as redshift decreases. The growing hot bubble is roughly
spherical by $z=1$, in agreement with the assumption of the analytic spherical
blast wave model in Chatterjee and Kosowsky (2007).

In order to further characterize this expanding hot bubble, Fig.~2 displays
maps of the difference between the two simulation with black hole modeling
and without, in the same 200 kpc regions of Fig.~1.  The top and second rows
show the most massive black hole at $z=3$ and $1$ respectively while the third
and forth row show the second black hole at the same redshifts.  In this
figure, the left column shows the logarithm of the $y$ distortion, the {\bf central} column is the logarithm of the mass-weighted temperature in
units of Kelvin and the right is the logarithm of projected electron number density in units of cm$^{-2}$,.  At $z=3$ a residual $y$ distortion is evident and concentrated around the black hole, with little effect further out; the peak
$y$ distortion due to the black hole is on the order of $10^{-7}$,
corresponding to an effective temperature shift of the order of 1 $\mu$K.  By
$z=1$, the energy injected into the center has propagated outwards, forming a
hot halo around the black hole.  Table 2 shows the respective black hole
accretion rates at different redshift for the two black holes in Figure 1 and
2.  It is evident that the highest amplitude of $y$ distortion is associated
with the most active, high-redshift epochs of accretion, when large amounts of
energy are coupled to the surrounding gas via the feedback process. At $z=1$
the black hole accretion rate has dropped so the $y$ distortion has a smaller
amplitude but has spread over a larger region (Fig. 2).

\subsection {Angular Profiles}
For the two black holes shown in Figure~1 we see an overall enhancement in the
SZ signal due to quasar feedback. This agrees with the
simulations in Scannapieco, Thacker \& Couchman 2008. To further quantify the effects of quasar feedback we average the SZ signal in annuli around the black hole and examine the angular profile of the resulting $y$ from the hot bubble in Figure 3 and 4.

Figure 3 shows the average angular profiles of the total $y$ distortion around
the two objects in the maps in Figure~1. The black dashed, blue dot-dashed and
red solid lines are for $z=1$, $z=2$, and $z=3$ respectively. In both cases
the $y$ increases with time between $\sim$ 10 to 25 arcsecond separation from
the black hole. $y$ gets steadily larger as the feedback energy spreads over
this volume (see also Fig.4). At $z=3$ the $y$ profile is steeper in the
central regions with a significant peak (in particular for the second quasar)
at scales below 5 arcseconds. The bumps in the profiles are due to
concentrations of hot gas or occasional other black holes which are included
in the total average signal.  $y$ typically reaches its highest central peaks
at time when the quasar is most active (the black hole accretion rate is high
- see Table 2), and hence large amounts of energy are coupled to the
surrounding gas according to our feedback prescription. For example, the $z=3$
curve in the right panel shows the black hole at a particularly active phase;
the central $y$ distortion corresponds to a temperature difference of over 4
$\mu$K.  At $z=2$ this central distortion is smaller by a factor of 20, while
it is larger by a factor of 10 at an angular separation of 10 arcseconds.
Figure 3 shows the total SZ effect in the direction of a quasar resulting from
the superposition of the SZ signature from quasar feedback plus the SZ
distortion from the rest of the line of sight due to the surrounding adiabatic
gas compression, which is expected to form an average background level in the
immediate vicinity of the back hole.

In order to clearly disentangle the contribution due to quasar feedback, in
Figure~4, we plot the fractional change in $y$ distortion between the
simulation with and without black hole modeling, at two different
redshifts. These are the profiles corresponding to the maps shown in
Figure~2. It is clear that the local SZ signature is largely dominated by the
energy output from the black hole, giving a factor between 300 to over 3000
(for the second black hole at $z=3$ in right panel) increase in $y$ near the
black hole. Our results are also consistent with the expected $y$ distortion
from the thermalized gas in the host halos containing these black holes (which
are on the order $10^{12}M_{\odot}$ to $10^{13}M_{\odot}$) and is the range
$10^{-9}$ to $10^{-7}$ (see also Komatsu \& Seljak 2002). The largest peak in $y$ distortion enhancement due to quasar feedback generally lies within 5
arcseconds of the black hole.

\subsection{Black Hole Mass Scaling Relations}
Since the SZ effect from the region around the black holes we analyzed in the
previous section is dominated by the quasar feedback, we investigate whether
a correlation between black hole mass and $y$ distortion exists for the
population as a whole (see also Colberg \& Di Matteo 2008 for other scaling
relations between $M_{BH}$ and host properties). The top row of Figure~5 plots
the mean $y$ distortion, computed over a sphere of radius 200 kpc/$h$
(i.e. the same as in the maps, corresponding to 20 arcseconds) versus black
hole mass for all black holes in the simulations with $M_{BH} >
10^{7}M_{\odot}$ at $z=1,2$ and $3$ (from right to left respectively). The
size of the region is chosen to sample the entire region of distortion due to
the quasar feedback, while minimizing bias from the local environment (Fig. 3
and 4). The mass cut-off is chosen to (a) minimize effects due to lack of
appropriate resolution in the simulations as well as (b) produce SZ
distortions that may be detectable by current or upcoming experiments.  

Simple power law fits to the $y$ distortion as a function of black hole mass
show a redshift evolution with the scaling becoming steeper with decreasing
redshift. Table 3 summarizes our results from the fits.  The trends show a
close correspondence between the mean $y$ parameter and the total feedback
energy as measured from $y$.  In order to further investigate the reason for
$y-M_{BH}$ relations, in the bottom row of Figure~5 we plot the accretion rate
versus black hole mass at redshifts 3, 2, and 1 for the same sample as in the
top panel and perform similar power-law fits (see Table 3). The trends in
accretion rate versus $M_{BH}$ are qualitatively similar to the top panel,
demonstrating the connection of the $y$ distortion due to quasar feedback with
the black hole accretion rate and black hole mass. In particular, at $z=1$ the
relations get steeper as expected if the largest fraction of black holes are
accreting according to the Bondi scaling (e.g., $\dot{m} \propto M_{BH}^2$)
and shallower with increasing redshift when most black holes are accreting
close to the critical Eddington value (e.g., $\dot{m} \propto M_{BH}$).  Of
course, the accretion rate depends not only on black hole mass but also on the
properties of the local gas and is also regulated by the large scale gas
infall driven by major mergers, which peak at higher redshifts (Di Matteo et
al. 2008). The ratio of the slopes (accretion rate to y distortion) for the fits shown in table 3 are 1.32, 0.65 and 0.73 at redshifts 3.0, 2.0 and 1.0 respectively. This shows the agreement of the top and the bottom panels in
Figure 5, and the close connection between accretion history and SZ
distortion: the SZ effect tracks closely quasar feedback and is promising
probe of black hole accretion. The largest amplitudes of SZ signal from
quasar is expected from $z\sim 2-3$ at a time close to the peak of the quasar
phase in galaxies.

\begin{figure*}
  
    \begin{tabular}{c}
      \resizebox{60mm}{!}{\includegraphics{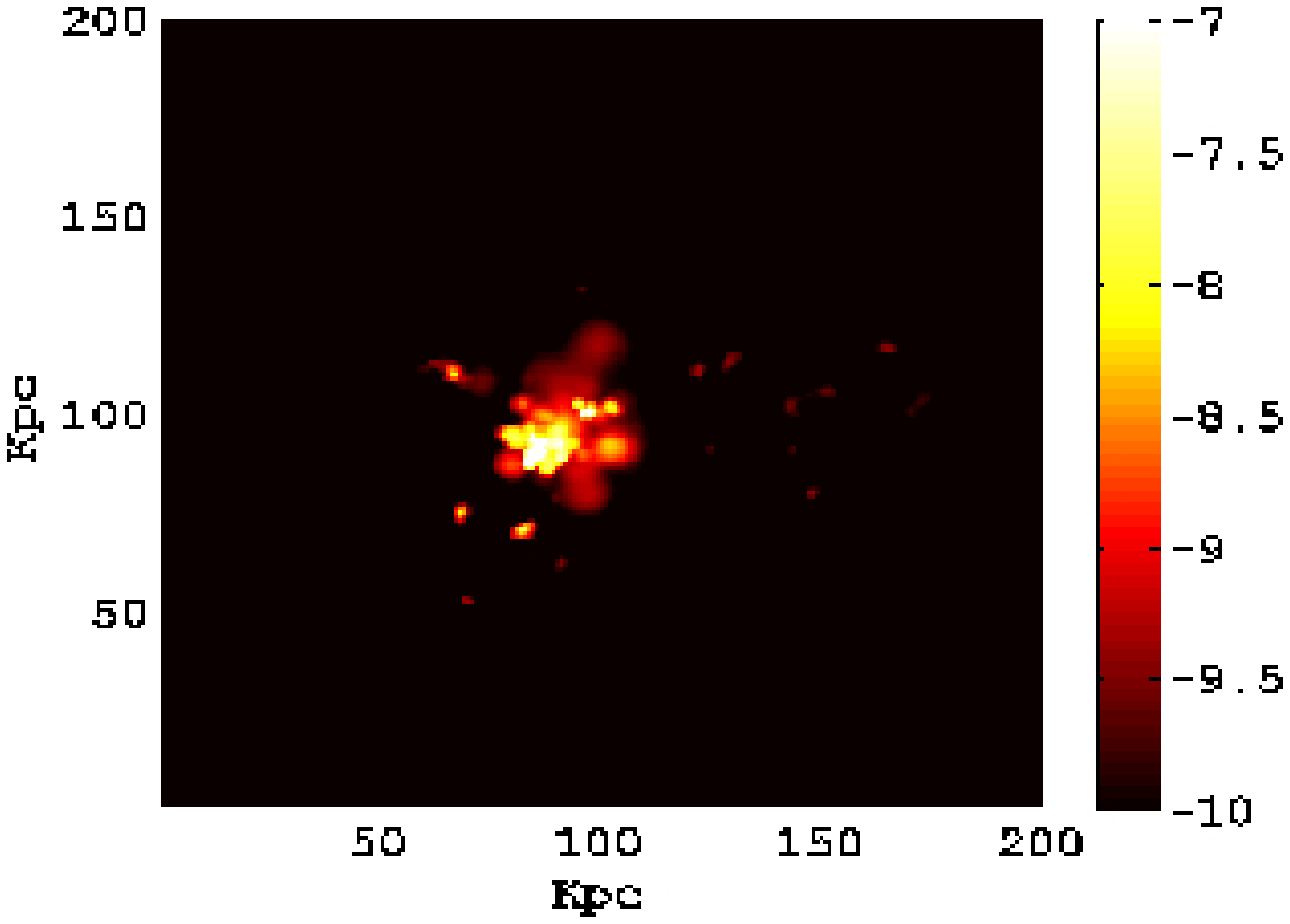}}
\resizebox{60mm}{!}{\includegraphics{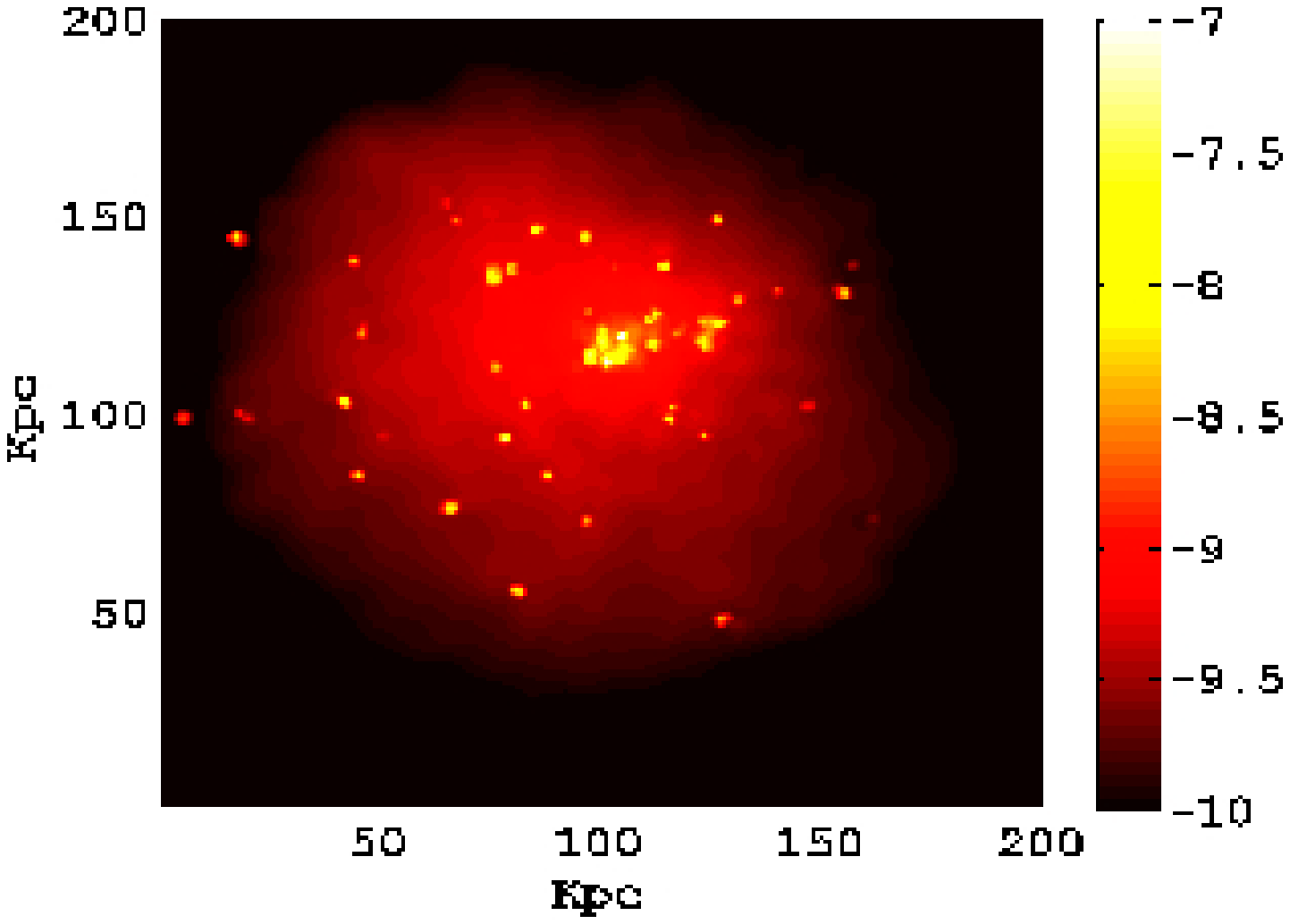}}
         \resizebox{60mm}{!}{\includegraphics{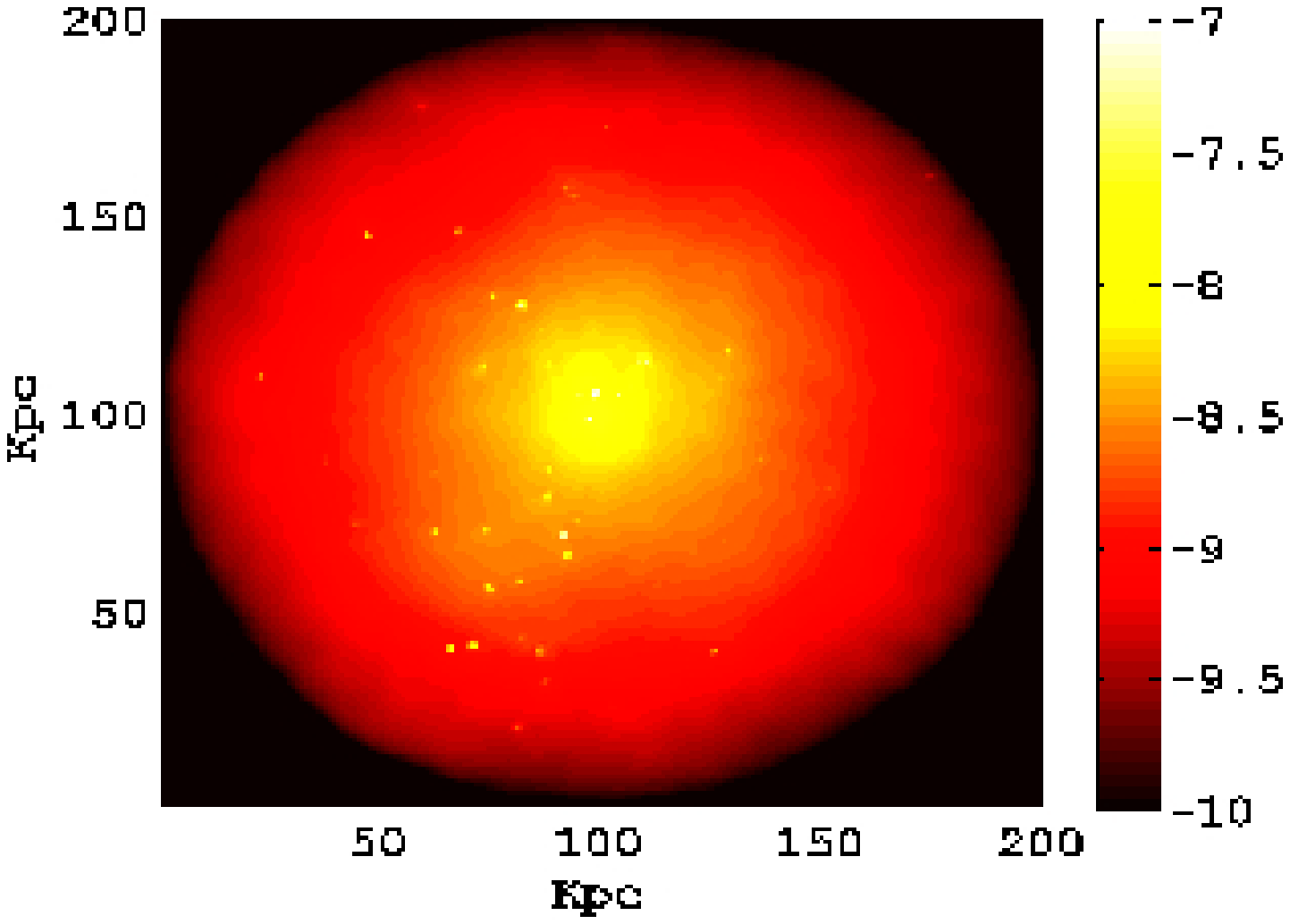}}\\
 \resizebox{60mm}{!}{\includegraphics{bh858100feed1.eps}}
\resizebox{60mm}{!}{\includegraphics{bh1115100feed1.eps}}
         \resizebox{60mm}{!}{\includegraphics{bh1266100feed1.eps}}\\

\end{tabular}
        \caption{The top row shows $y$-distortion maps of the most massive
        black hole at $z=3$ (left), $z=2$ (center) and $z=1$ (right) in a
        higher-resolution (D6) simulation. The bottom row shows the same thing
        for a lower-resolution (D4) run. The black hole masses in the two
        simulations are slightly different from each other: for $z=1$ masses
        $4.29 \times 10^{9} M_{\odot}$ (D4) and $2.96 \times 10^{9} M_{\odot}$
        (D6), for $z=2$ masses $2.76 \times 10^{9} M_{\odot}$ (D4) and $1.83
        \times 10^{9} M_{\odot}$ (D6), and for $z=3$ masses $7.35 \times
        10^{8}M_{\odot}$ (D4) and $8.56 \times 10^{8}M_{\odot}$ (D6). The box
        size is 200 kpc square. The difference in peak $y$ value for D4 and D6
        varies from 22\% ($z=3$) to 6\% ($z=2$) and it is higher for the
        higher-resolution simulation at all three redshifts.}
  \end{figure*}

\subsection{Resolution Test}
 In the previous section we have made use of the D4 (Table 1) simulations from
 our analysis. At this resolution we have two identical realizations, with and
 without black hole modeling, allowing us to carry out detailed comparisons of
 the effects of the quasar feedback. We now wish to assess possible effects
 due to numerical resolution by making use of the D6 (BHCosmo) run (see also
 Di Matteo et al. 2008, Croft et al. 2008 and Bhattacharya DiMatteo \& Kosowsky
 2008 for additional resolution studies).  Figure 6 shows the $y$ distortion maps for the most massive black hole at redshifts 3, 2, and 1. The top row is for the higher-resolution BHCosmo run and the bottom row is for the lower-resolution run (D4). Our results at the lower resolution appear reasonably well
 converged, though with some differences. The central black hole masses in the
 two runs differ somewhat. At $z=1$, $2$, and $3$, the black hole masses in the D4 and BHCosmo run are ($4.29 \times 10^{9}M_{\odot}$, $2.96 \times 10^{9}
 M_{\odot}$), ($2.76 \times 10^{9} M_{\odot}$, $1.85 \times 10^{9} M_{\odot}$)
 and ($7.35 \times 10^{8}M_{\odot}$, $8.56 \times 10^{8}M_{\odot}$)
 respectively. It clear that the difference in resolution is affecting the
 black hole mass as expected from modest changes in mass accretion rate (which
 is sensitive to the gas properties close to the black hole). Also,
 more small scale structure in the gas distribution is evident at higher
 resolution, as expected. This  affects the amplitude of the total SZ flux
 which is enhanced by about 6\% at $z=2$ and by about 22\% at $z=3$ (when it
 is most peaked around the black hole) in the higher resolution run.

\section{Detectability and Discussion} 
Observationally, quasar feedback is directly detectable by resolving
Sunyaev-Zeldovich peaks on small angular scales of tens of arcseconds with
amplitudes of up to a few $\mu$K above the immediately surrounding region. The
combination of angular scale and small amplitude make detecting this effect
very challenging, at the margins of currently planned experiments.  The
necessary sensitivity requires large collecting areas, while the angular
resolution needed points to an interferometer in a compact configuration, or a
large single-dish experiment.  Since the SZ signal is manifested as a peak
over the surrounding background level, a region substantially larger than the
SZ peak must be imaged. This requires a telescope having sufficient resolution to resolve the central peak in the SZ distortion in an SZ image and enough field of view so that the peak could be identified. An example is the compact ALMA subarray known as the
Atacama Compact Array (ACA), composed of 12 7-meter dishes. The ALMA
sensitivity calculator gives that the synthesized beam for this array is about
14 arcseconds, and the integration time required to attain 1 $\mu$K
sensitivity per beam at a frequency of 145 GHz and a maximum band width of 16
GHz is on the order of 1000 hours (ALMA sensitivity calculator).  A very deep survey with this instrument could detect the SZ effect from individual black holes. The 50-meter Large Millimeter-Wave Telescope instrumented with the AzTEC bolometer array detector will have a somewhat similar sensitivity but detectibility would require a very deep (thousands of hours) integration time. The Cornell-Caltech Atacama Telescope (CCAT), a 25-meter telescope, estimates a possible pixel sensitivity for SZ detection at 150 GHz of 310 $\mu{\rm K}\,{\rm s}^{1/2}$ for 26 arcsecond pixels, so a 30 hour observation could give 1$\mu$K pixel noise. These pixels would not be small enough to resolve the hot halo around a black hole, but might be able to detect the difference in a single pixel due to black hole emission compared to the surrounding pixels. Aside from raw sensitivity and angular resolution, a serious difficulty with direct detection is the confusion limit from infrared point source emission; these sources are generally high-redshift star forming galaxies with a high dust emission. CCAT estimates show that their one-source-per-beam confusion limit will be around 6 $\mu$K at 150 GHz (Golwala 2006). This will present substantial difficulties for detecting a 1 $\mu$K temperature distortion if accurate. It is noted that the observations in the sub-millimeter band is limited by confusion noise and so another possibility of direct detection of the signal through radio frequency telescopes could be considered. Massardi et al. 2008 shows that the confusion due to dusty galaxies is lower at 10 GHz then at 100 GHz. The authors show that for galactic scale SZ effect the optimal frequency range for detection is between 10 to 35 GHz. However substantial confusion from radio galaxies at these low frequency observations would still be a challenging issue in the direct detection of the signal.

Given these substantial difficulties associated with direct detection, an
alternate route may be necessary. Cross-correlation of arcminute-resolution
microwave maps with optically selected quasar or massive galaxies is a second
possible detection strategy (Chatterjee and Kosowsky 2007, Scannapieco,
Thacker, and Couchman 2008). By averaging over large numbers of objects, we
can have an estimate of a small mean black hole distortion signal from the
noise in the maps.  The primary challenge with this technique is the direct
emission from quasar in the microwave band. It may be possible to select a
sample of quasar which is sufficiently radio-quiet that the cross-correlation
is not dominated by the intrinsic emission. Another possibility is to select
massive field galaxies under the assumption that they harbor a central massive
black hole which at one time was active; the hot bubble produced has a cooling
time comparable to the Hubble time, so formerly active galaxies should still
have an SZ signature. Finally, the SZ effect from black holes is in addition
to the SZ emission from any hot gas in which the black hole's host galaxy is
embedded. Massive galaxies trace large-scale structure, and any
cross-correlation will also detect this signal. Although the observational requirements for the cross-correlation method are plausible the scopes for detectibility with this method is still limited by confusion noise. Stacking microwave (SZ) maps in the direction of known quasars would also serve as an independent route in detecting the signal (Chatterjee \& Kosowsky 2007). This can improve the signal to noise by a substantial amount although this method would still be limited by the uncertainties described above. Quantifying in detail the observable signal (which will need to be
disentangled from other confusions such as dusty galaxies, radio galaxies etc.) for the possible direct
detections methods or from cross-correlation analysis that we have
discussed is beyond the scope of this paper and we defer it to a future work. The
simulations and maps presented here provide a basis for further modeling of
all these effects. 

The main conclusions drawn from this work are summarized as follows. We have used the first cosmological simulations to incorporate realistic black hole growth and feedback to produce simulated maps of the Sunyaev-Zeldovich
distortion of the microwave background due to the feedback energy from
accretion onto supermassive black holes. These simulations address the rapid
accretion phases of black holes: periods of strong emission are typically
short-lived and require galaxy mergers to produce strong gravitational tidal
forcing necessary for sufficient nuclear gas inflow rates (Hopkins, Narayan \&
Hernquist 2006; Di Matteo et al. 2008). The result is heating of the gas
surrounding the black hole, so that the largest black holes produce a
surrounding hot region which induces a $y$-distortion (related to a temperature distortion) with a characteristic
amplitude of a few $\mu$K. We have obtained a scaling relation between the black hole mass and their SZ temperature decrement, which in turn is a measure of the amount of feedback energy output. The correspondence between the y distortion and the accretion rates is not exact but there is a close association which shows the correlation between feedback output and black hole activity. From our results we have shown that with the turn on of AGN feedback the signal gets enhanced largely and the enhancement is predominant at angular scales of 5 arcseconds. Finally we have shown that there is a fair probability of detecting this signal even from the planned sub millimeter missions.  

\par The role of energy feedback from quasars and from star
formation is known to have substantial impact on the process of galaxy
formation and evolution of the intergalactic medium, but the details of this
process are not well understood. Probes based on Sunyaev-Zeldovich distortions
are challenging, but an eventual detection can be used to put useful constraints and checks on models of AGN feedback.

\section{acknowledgments}
SC would like to thank Bruce Partridge and James Moran for helpful discussions
on experimental capabilities of various telescopes. Special thanks to Mark
Gurwell for helping with the sensitivity calculation for SMA. SC and AK would
also like to thank Christoph Pfrommer for some initial discussions on the
project. Thanks to Jonathan Las Fargeas who helped with the analysis,
supported by NSF grant 0649184 to the University of Pittsburgh REU
program. We would also like to thank the referee for valuable suggestions on improvement of the paper.This work was supported at the University of Pittsburgh by the
National Science Foundation through grant AST-0408698 to the ACT project, and
by grant AST-0546035. At CMU this work has been supported in part
through NSF AST 06-07819 and NSF OCI 0749212. SC was also partly funded by the Zaccheus Daniel Fellowship at the University of Pittsburgh.

 \end{document}